\documentclass[a4paper,12pt]{article}
\usepackage{jheppub,esint,shuffle,psfrag}
 \usepackage[utf8]{inputenc}
 
\usepackage{tikz}
\usetikzlibrary{arrows.meta,decorations.markings}

\usepackage{esint} 
\usepackage{breqn}
\usepackage{bm}
\def \tr {\mathop{\rm tr}\nolimits}
\def \Im {\mathop{\rm Im}\nolimits}
\def \Re {\mathop{\rm Re}\nolimits}

\newcommand\lr[1]{{\left({#1}\right)}}
\newcommand \widebar [1] {\overline{#1}}

\newcommand \vev [1] {\langle{#1}\rangle}
\newcommand \VEV [1] {\left\langle{#1}\right\rangle}
\newcommand \ket [1] {|{#1}\rangle}
\newcommand \bra [1] {\langle {#1}|}
\newcommand\re[1]{(\ref{#1})}

\def \qqqquad {\qquad\qquad}

\newcommand{\ft}[2]{{\textstyle\frac{#1}{#2}}}

\def\numberbysection{\@addtoreset{equation}{section}
                     \def\theequation{\thesection.\arabic{equation}}}
 
\begin{document}

\vspace*{0cm }

\author{Bercel Boldis$^{a,b}$, Gregory P. Korchemsky$^{c}$  and  Alessandro Testa$^{c}$  }
\affiliation{
$\null$
$^a$Department of Theoretical Physics, Institute of Physics, Budapest University of Technology and Economics M\H{u}egyetem rkp. 3, 1111 Budapest, Hungary
\\
$\null$ 
$^b$HUN-REN Wigner Research Centre for Physics, Konkoly-Thege Miklos ut 29-33, 1121 Budapest, Hungary
\\		
$\null$
$^c${Institut de Physique Th\'eorique\footnote{Unit\'e Mixte de Recherche 3681 du CNRS}, Universit\'e Paris Saclay, CNRS,  91191 Gif-sur-Yvette, France}   
}
\title{Wilson loops in ABJM theory reloaded}
\abstract{\small
We present a new technique for computing supersymmetric Wilson loops in the ABJM theory via supersymmetric localization, valid for arbitrary values of the rank of the gauge group  $N$  and the Chern--Simons level $k$. The approach relies on an operator representation of the Wilson loops within the Fermi gas formalism in terms of the resolvent of a certain integral operator previously encountered in the computation of the ABJM partition function on the round three-sphere. By deriving a set of nontrivial relations for this resolvent, we obtain exact expressions for the generating functions of Wilson loops in terms of the partition function. For large $k$, these expressions reproduce the weak-coupling expansion of the Wilson loops, and in the large-$N$ limit at fixed $k$ they match previously obtained high-precision numerical results. This analysis also resolves the longstanding discrepancy between numerical data and the semiclassical expression for the $1/6$ BPS Wilson loop.
}

\maketitle
  
\section{Introduction and summary}

Supersymmetric gauge theories with extended supersymmetry provide a powerful framework for studying nonperturbative aspects of quantum field theory. In three dimensions, a  prominent example is the ABJM model~\cite{Aharony:2008gk,Aharony:2008ug}.   It describes two supersymmetric Chern--Simons theories with gauge group $ U(N) $ and opposite levels, $ k $ and $ -k $, coupled to four matter supermultiplets in the bifundamental representation of $ U(N) \times U(N) $.
 The theory exhibits $ \mathcal{N} = 6 $ superconformal symmetry which, for the special values $ k = 1 $ and $ k = 2 $, is enhanced to $ \mathcal{N} = 8 $, the maximal amount of supersymmetry allowed in three dimensions. 

The parameter space of the ABJM model is fully characterised by two independent positive integers: the rank $ N $ of the gauge group and the Chern--Simons level $ k $. The inverse level $ 1/k $ defines the coupling constant of the theory. The weak and strong coupling regime of the theory correspond, respectively, to the large and small $k$ limit.

At large $N$, two qualitatively different regimes emerge, depending on the hierarchy between $ N $ and $ k $. When both parameters are large, observables can be systematically expanded in powers of $ 1/N^2 $ by keeping the effective 't~Hooft coupling $ \lambda = N/k $ fixed. Within the AdS/CFT correspondence \cite{Aharony:1999ti}, this limit corresponds to the perturbative genus expansion of a  type~IIA superstring propagating on $ \mathrm{AdS}_4 \times \mathbb{CP}^3 $~\cite{Benna:2008zy,Arutyunov:2008if,Stefanski:2008ik,Hosomichi:2008jb}.

The second regime arises when the level $ k $ is kept fixed as $ N \to \infty $ or, equivalently, when $ \lambda $ grows linearly with $ N $. In this regime, the curvature of the type~IIA background becomes large and the ten-dimensional description is no longer valid. The appropriate dual description is instead provided by an eleven-dimensional M-theory compactified on $ \mathrm{AdS}_4 \times S^7/\mathbb{Z}_k $. This  regime provides a nonperturbative completion of the ABJM theory and plays a pivotal role in understanding the dynamics of multiple M2-branes.

Supersymmetric localization~\cite{Pestun:2016zxk} provides a powerful framework for studying different regimes of the ABJM theory. For a special class of observables preserving supersymmetry, this technique  allows us to reduce the infinite-dimensional path integral to a finite-dimensional matrix integral. Originally developed for four-dimensional supersymmetric Yang-Mills theories~\cite{Pestun:2007rz}, it was later generalised to three-dimensional supersymmetric Chern--Simons theories~\cite{Kapustin:2009kz}. 

In the ABJM theory, the matrix model obtained via supersymmetric localization provides a nonperturbative representation of a wide class of observables, including the partition function on the three-sphere $ S^3 $, circular Wilson loops~\cite{Drukker:2008zx,Chen:2008bp,Drukker:2009hy,Drukker:2019bev} and the bremsstrahlung function~\cite{Lewkowycz:2013laa,Bianchi:2014laa,Bianchi:2017ozk,Bianchi:2018bke,Bianchi:2018scb}. 
Solving this matrix model and deriving exact expressions for the observables that are valid for arbitrary values of parameters of the ABJM theory is an extremely difficult and challenging task. 

A powerful technique for addressing this problem is the \emph{Fermi gas} approach~\cite{Marino:2011eh}. In this framework, the ABJM matrix model is reinterpreted as a one-dimensional quantum gas of interacting fermions, with the Chern--Simons level $ k $ playing the role of an effective Planck constant, $ \hbar = 2\pi k $.
This approach  has been successfully applied to the computation of the partition function $ Z(N,k) $ of the ABJM theory on the round three-sphere $ S^{3} $, see \cite{Hatsuda:2015gca,Marino:2016new} for reviews. 
Remarkably, the generating function $\Xi(z,k)= \sum_{N\ge 0} z^{N} Z(N,k) $, which has the meaning of the grand canonical partition function of the Fermi gas, admits a compact representation as a Fredholm determinant of a certain integral operator.~\footnote{It is noteworthy that various observables in four-dimensional maximally supersymmetric Yang-Mills theory can likewise be expressed in terms of a Fredholm determinant of an integrable Bessel kernel (see \cite{Beccaria:2022ypy} and references therein). This representation is particularly powerful, as it enables the computation of these observables for arbitrary values of the 't Hooft coupling \cite{Bajnok:2024epf,Bajnok:2024ymr,Bajnok:2024bqr}.} In the strong-coupling regime, corresponding to the limit $ \hbar = 2\pi k \to 0 $, the Fredholm determinant can be evaluated using a semiclassical expansion. When supplemented with nonperturbative (instanton) corrections, this analysis leads to the expression for the partition function in terms of Airy function.

The Fermi gas formalism can also be applied to the computation of supersymmetric Wilson loops. In the ABJM theory, one can define two distinct supersymmetric Wilson loops, $W^{1/6}$ and $W^{1/2}$, which preserve four and twelve superconformal symmetries, respectively, out of the total $24$ of the theory  \cite{Drukker:2008zx,Chen:2008bp,Drukker:2009hy,Drukker:2019bev}. The $1/6$~BPS Wilson loop is defined as the vacuum expectation value
\begin{align}\label{W-1/6}
W_n^{1/6} 
= \frac{1}{N}\,
\biggl\langle  
\, 
\tr_{\Box}\, 
P \exp\!\left[
\int_{0}^{2\pi n}\! ds \left(
i A_\mu(x)\,\dot x^\mu 
+ \frac{2\pi}{k}\, |\dot x|\, M^{I}_{J}\, C_I(x)\, \bar C^{J}(x)
\right)\right] 
\biggr\rangle ,
\end{align}
where the contour $x_\mu = x_\mu(s)= x_\mu(s+2\pi)$ winds $n$ times around the unit circle in $\mathbb R^3$, and $\dot x^\mu = \partial_s x^\mu(s)$ denotes the tangent vector.  
The exponent in \re{W-1/6} contains the gauge field $A_\mu(x)$ of the Chern--Simons theory with level $k$~\footnote{An analogous Wilson loop can be defined for the gauge field of the Chern--Simons theory at level $(-k)$. It is related to~\re{W-1/6} by complex conjugation, or equivalently by the replacement $k \to -k$.} as well as the scalar matter fields $C_I$ (with $I=1,\dots,4$), which transform in the bi-fundamental representation of the gauge group $U(N)\times U(N)$ and in the fundamental representation of the $R$-symmetry group $SU(4)$.  
The constant matrix $M^{I}_{J}$ is fixed by supersymmetry, it can be brought to the diagonal form $M = \mathrm{diag}(1,1,-1,-1)$ by an $SU(4)$ transformation.  
The trace in~\re{W-1/6} is taken in the fundamental representation of the gauge group.
 
The $1/2$~BPS Wilson loop is defined in terms of the holonomy of the supergroup $U(N|N)$. Its construction involves the gauge fields of both Chern--Simons theories, together with the matter fields (scalars and fermions). The expectation value of the $1/2$~BPS Wilson loop with winding number $n$ can be expressed as a linear combination of the $1/6$~BPS Wilson loops~\re{W-1/6} associated with the Chern--Simons theories at levels $k$ and $-k$,
\begin{align}\label{W-1/2}
W_n^{1/2} = W_n^{1/6} - (-1)^n\, \widebar{W_n^{1/6}}\,.
\end{align}
Both Wilson loops~\re{W-1/6} and~\re{W-1/2} depend on a nonnegative integer $n$, which specifies the winding number.  
For $n=0$, they satisfy the normalization condition $W_{0}^{1/6}=1$ and similarly $W_{0}^{1/2}=0$.
 As a consequence of \re{W-1/2}, it is sufficient to focus on the computation of the $1/6$~BPS Wilson loop~\re{W-1/6}.

The semiclassical expansion of the Wilson loops \re{W-1/6} and \re{W-1/2} within the Fermi gas approach was derived in \cite{Klemm:2012ii}. Neglecting nonperturbative (instanton) corrections, the resulting expressions for $W_n^{1/6}$ and $W_n^{1/2}$ take the form of ratios of Airy functions and their derivatives. In parallel development, the $1/6$ BPS Wilson loop was computed numerically with high precision for a wide range of parameters $k$ and $N$ in~\cite{Okuyama:2016deu}. These numerical studies revealed that, for winding number $n \ge 2$, the perturbative part of the $1/6$ Wilson loop differs from the semiclassical expression obtained in \cite{Klemm:2012ii}. This discrepancy was unexpected, especially given that the semiclassical expression for the partition function $Z(N,k)$ agrees perfectly with numerical data~\cite{Hatsuda:2012hm}. The origin of the mismatch remained unexplained.

The goal of this work is to develop a new technique for computing the supersymmetric Wilson loops \re{W-1/6} and \re{W-1/2} for arbitrary values of the parameters, without relying on the semiclassical expansion. 

In the Fermi-gas framework, the Wilson loop \re{W-1/6} admits an exact representation in terms of the resolvent of the same operator that appears in the Fredholm determinant expression for the partition function $ Z(N,k) $. The corresponding Fredholm determinant is known to satisfy a system of coupled nonlinear TBA-like equations. These equations were originally introduced independently in the study of $ \mathcal{N}=2 $ models in two dimensions~\cite{Cecotti:1992qh} and of self-avoiding 2d polymers~\cite{Zamolodchikov:1994uw}, and were subsequently proven in~\cite{Tracy:1995ax}.

Building on the method developed by Tracy and Widom in~\cite{Tracy:1995ax}, we derive a set of exact equations governing the function $ W_n^{1/6} $. We show that these equations are sufficiently powerful to determine the Wilson loops for arbitrary values of the parameters, without relying on the semiclassical approximation.

We show that for arbitrary $k$ the resulting expressions for the generating functions $\mathcal W^{1/6}_n(z)$ and
$\mathcal W^{1/2}_n(z)$ of the Wilson loops \re{W-1/6} and \re{W-1/2} admit a closed form representation in terms of the free energy $\mathcal F(z)$ (see \re{W-cf} below). This relation takes a remarkably simple form for $n=1$ and $k=4$
\begin{align}\label{W-ex1} 
\mathcal W^{1/2}_{n=1}(z)=\frac{1}{2}\sqrt{z-4+4\, e^{-2\mathcal F(z)}}\,.
\end{align}
Moreover, for arbitrary $k>4$ and $n=1,2$ the corresponding generating functions are related to each other as
\begin{align}\label{W-ex2} 
\mathcal W^{1/2}_{n=2}(z)=i\tan\!\Big(\frac{2\pi }{k}\Big) \Big[\Big(\mathcal W^{1/2}_{n=1}(z)\Big)^2+1-e^{-2\mathcal F(z)}\Big]\,.
\end{align}

We verified that the obtained results for the Wilson loops are in perfect agreement with the expressions obtained in \cite{Okuyama:2016deu} from numerical fitting. Combined with the explicit expressions for the free energy, they provide an efficient method for computing the Wilson loops for arbitrary values of the parameters of the ABJM theory.

The paper is organized as follows. In Section~\ref{sect2}, we review the Fermi gas approach and introduce the operator definition of the generating function $\mathcal W_n$ for the $1/6$ BPS Wilson loop. In Section~\ref{sect3}, we apply the Tracy--Widom approach to obtain an integral representation of $\mathcal W_n$ in terms of an auxiliary $\psi$--function, and derive a set of functional relations satisfied by this function. These relations are analyzed in Section~\ref{sect4}. In Section~\ref{sect5}, we combine the results of the preceding sections to compute the generating function $\mathcal W_n$. In Section~\ref{sect7}, we study the large-$z$ behavior of $\mathcal W_n$ and compare it with the corresponding expressions obtained from the semiclassical expansion and numerical fitting. Finally, Section~\ref{sect8} contains our concluding remarks. Additional technical details are collected in the appendices.

\section{Fermi gas approach}\label{sect2}
Supersymmetric localization  allows us to express the partition function of the ABJM theory on the three-sphere and the $1/6$ BPS Wilson loop \re{W-1/6} in terms of the same function $W_n=W_n(N,k)$.
This function is given by a finite-dimensional integral \cite{Kapustin:2009kz}
\begin{align}\label{Z}
W_n &= \frac{1}{(N!)^2} \int_{-\infty}^\infty \prod_{i=1}^N {d\mu_i d\nu_i\over (2\pi)^2} 
{\prod_{i<j} \left[2 \sinh \left( \frac{\mu_i - \mu_j}{2} \right) \right]^2 \left[2 \sinh \left( \frac{\nu_i - \nu_j}{2} \right) \right]^2 \over \prod_{i,j} \left[2 \cosh \left( \frac{\mu_i - \nu_j}{2} \right) \right]^2}
e^{ \frac{ik}{4\pi} \sum_{i=1}^N (\mu_i^2 - \nu_i^2)}\sum_{i=1}^N e^{n \mu_i} \,.
\end{align}
It depends on the parameters of the theory, namely the Chern--Simons level~$k$, the rank of the gauge group~$N$, and the winding number~$n$.

For arbitrary values of the parameters, the partition function $Z(N,k)$ and the $1/6$ BPS Wilson loop $W_n^{1/6}$ are given by
\begin{align}\label{Z-W}
Z  = {1\over N} W_0(N,k)\,,\qqqquad 
W_n^{1/6}= {W_n(N,k)\over W_0(N,k)}\,.
\end{align}
The factor of $1/N$ in the first relation is introduced to compensate the last factor on the right-hand side of \re{Z} evaluated for $n=0$.  The function $W_n(N,k)$
 differs from the $1/6$~BPS Wilson loop~$W_n^{1/6}$ by the overall factor~$W_0(N,k)$
and is therefore conventionally referred to as the  un-normalized Wilson loop.

For $n\ge 1$ the last factor on the right-hand side of~\re{Z} grows exponentially for large values of~$\mu_i$. As a result, the convergence of the integral \re{Z} is not guaranteed for arbitrary values of the parameters. A closer examination shows that this integral is well-defined only if the Chern--Simons level satisfies
\begin{align}\label{k>2n}
k > 2n\,.
\end{align}

To illustrate this, it is sufficient to consider the Abelian case with~$N=1$. 
In this case, the integral in \re{Z} can be evaluated explicitly, yielding \cite{Okuyama:2016deu}
\begin{align}\label{W-N=1}
W_n(N=1,k) = \frac{e^{i\pi n^2/k}}{4k\,\cos^2(\pi n / k)}\,.
\end{align}
This expression is regular for~$k$ obeying~\re{k>2n}, but it develops a singularity at~$k = 2n$. 
The singularity originates from the divergence of the integral~\re{Z} at this value of~$k$. 
As we demonstrate below, the same singular behavior persists for arbitrary~$N$.

\subsection{Generating functions}

To compute the functions \re{Z-W} for arbitrary $N$ it is convenient to introduce an auxiliary fugacity parameter $z$ and define generating functions
\begin{align}\notag\label{Xi}
{}& \Xi(z,k) = 1+ \sum_{N\ge 1} z^N Z(N,k)\,,
\\
{}& \mathcal W_n(z,k) = {1\over \Xi(z,k)}\sum_{N\ge 1} z^N W_n(N,k)\,.
\end{align}
As follows from the first relation in \re{Z-W} these two functions are not independent and are related to each other as
\begin{align}\label{W_0}
\mathcal W_0(z,k) = z\partial_z \log \Xi(z,k) \,.
\end{align}
The advantage of introducing the generating functions \re{Xi} is that the function $\Xi(z,k)$ can be interpreted as the grand canonical partition function of the Fermi gas with the chemical potential $\mu$ related to the fugacity parameter as
\begin{align}\label{z-mu}
z=e^\mu\,.
\end{align}
In a similar manner, the generating function $\mathcal W_n(z,k)$ has the meaning of the expectation value of the holonomy $\tr(U^n)=\sum_{i=1}^N e^{n \mu_i}$ in a grand canonical ensemble. 

Having computed the generating functions \re{det1} and \re{det3} for arbitrary $z$ we can apply \re{Xi}  to find the partition function $Z(N,k)$ and un-normalized Wilson loop $W_n(N,k)$ as
\begin{align}\notag\label{Z,W}
{}& Z(N,k) = \int_{-i\pi}^{i\pi} {d\mu\over 2\pi i} e^{-N\mu} \, \Xi(e^\mu,k)\,, 
\\
{}& W_n(N,k) = \int_{-i\pi}^{i\pi} {d\mu\over 2\pi i} e^{-N\mu} \, \Xi(e^\mu,k) \mathcal W_n(e^\mu,k)\,.
\end{align} 

Applying \re{Z-W} and replacing the functions $Z(N,k)$ and $W_n(N,k)$ with their integral representation \re{Z}, we obtain rather complicated expressions for the generating functions \re{Xi}. As was shown in \cite{Marino:2011eh}, these expressions can be cast into a compact form by noticing that the function $Z(N,k)$ admits a determinant representation
\begin{align} 
{}& Z (N,k)=  \frac{1}{N!} \int d^Nx \, \det\| \rho(x_i,x_j)\|_{_{1\le i,j\le N}}\,,
\end{align}
where the integration measure is  $d^N x= dx_1\dots dx_N$ and the function $\rho(x,y)$ is defined as
\begin{align}\label{rho-fun}
 \rho(x,y) = {1\over 8\pi k  \cosh(\frac{x}2) \cosh({x-y\over 2k})}\,.
\end{align}
The dependence of $Z (N,k)$ on the Chern-Simons level $k$ resides in this function.

Substituting the above relations into \re{Xi} leads to the representation of the generating function $\Xi(z,k)$ as a Fredholm determinant,  
\begin{align}\label{det1}
\Xi(z,k) = \det(1 + z\, \boldsymbol{\rho})\,,
\end{align}
where $\boldsymbol{\rho}$ is an integral operator with kernel $\langle x | \boldsymbol{\rho} | y \rangle = \rho(x,y)$.  
This operator admits a quantum mechani\-cal realization,
\begin{align}\label{rho-oper}
\boldsymbol{\rho} = \frac{1}{2\cosh(\boldsymbol{x}/2)} \, \frac{1}{2\cosh(\boldsymbol{p}/2)}\,,
\end{align}
where the operators $\boldsymbol{x}$ and $\boldsymbol{p}$ satisfy the canonical commutation relation
$
[\boldsymbol{x}, \boldsymbol{p}] = 2\pi i k 
$.

In a similar way, the generating function $\mathcal{W}_n(z,k)$ defined in \re{Xi} can be written in the operator form \cite{Klemm:2012ii,Hatsuda:2013yua},
\begin{align}\label{det2}
\mathcal{W}_n(z,k) = \tr\left( \frac{z\, \boldsymbol{\rho}}{1 + z\, \boldsymbol{\rho}} \, \boldsymbol{U}^n \right) \equiv   \int_{-\infty}^\infty dx\,\VEV{x\left| \frac{z\, \boldsymbol{\rho}}{1 + z\, \boldsymbol{\rho}}\boldsymbol{U}^n\right|x},
\end{align}
where the operator $\boldsymbol{U} = e^{(\boldsymbol{x} + \boldsymbol{p})/k}$
provides a representation of the holonomy in the grand canonical ensemble. We verify that for $n=0$ the relations \re{det1} and \re{det2} are in agreement with \re{W_0}. For $n\neq 0$, the operator in \re{det2} can be simplified using the Baker-Campbell-Hausdorff formula as $\bm U^n = e^{n\bm p\over k}e^{n(\bm x+i\pi n)\over k}$ leading to the following representation \cite{Okuyama:2016deu}
\begin{align}\label{det3}
\mathcal{W}_n(z,k) = \int_{-\infty}^\infty dx\, e^{n( x+i\pi n)\over k}\VEV{x\left| \frac{z\, \boldsymbol{\rho}}{1 + z\, \boldsymbol{\rho}}\right|x+2\pi i n}.
\end{align}
Note that the matrix element on the right-hand side involves an argument shifted into the complex plane. This shift is generated by the operator $e^{n\boldsymbol{p}/k}$.

As we demonstrate below, relations \re{det1} and \re{det3} provide a powerful tool for evaluating un-normalized Wilson loop $W_n(N,k)$ for arbitrary positive $N$ and $k$. For the first few values of $N$, this function can be found from small $z$ expansion of the generating functions \re{Xi}. Expanding the right-hand side of  \re{det1} and \re{det3} in powers of $z$, we get from \re{Xi}  
\begin{align} \label{W-small-z}
\sum_{N\ge 1} z^N W_n(N,k) {}&= z\tr(\bm\rho\, \bm U^n)+z^2 \Big[\tr(\bm\rho\, \bm U^n)\tr \bm\rho - \tr(\bm\rho^2\, \bm U^n)\Big]+ O(z^3)\,.
\end{align}
Matching the coefficients on both sides, we can obtain expressions for $W_n(N,k)$ for $N=1,2,\dots$.

For instance, for $N=1$ we find
\begin{align}
W_n(N=1,k) = \tr(\bm\rho\, \bm U^n) =    \int_{-\infty}^\infty dx\, e^{n( x+i\pi n)\over k} \rho(x,x+2\pi i n)\,.
\end{align}
Replacing the function $\rho(x,y)$ with its expression \re{rho-fun}, we correctly reproduce \re{W-N=1}. For $N\ge 2$, the corresponding expressions for $W_n(N,k)$ are given by multiple integrals which cannot be computed in a closed form for general $k$.
 
To obtain $W_n(N,k)$ for large $N$ we have to determine the generating functions \re{det1} and \re{det3} at large $z$. We solve this problem in the next section.
 
\subsection{Semiclassical approximation}  

In the quantum mechanical formulation of the generating functions \re{det1} and \re{rho-oper}, the parameter $k$ determines the Planck constant, $\hbar = 2\pi k$. Recall that this parameter also specifies the Chern-Simons level and therefore takes only positive integer values. Nevertheless, it proves convenient to relax this restriction and examine the small-$k$ expansion of the generating functions \re{det1} and \re{det3}.

The advantage of considering this regime is that, within the Fermi gas framework, the small-$k$ behavior of the generating functions $\Xi(z,k)$ and $\mathcal{W}_n(z,k)$ defined in \re{Xi} can be systematically derived from the semiclassical WKB expansion of the quantum Fermi gas.  
It is important to emphasize, however, that it is not obvious \emph{a priori} whether this expansion remains valid for finite values of $k$, or, equivalently, whether the semiclassical expressions can be analytically continued from small $k$ to positive integer values without encountering singularities.  
This question has been investigated in \cite{Hatsuda:2012hm,Okuyama:2016deu} by comparing semiclassical predictions with numerical evaluations of \re{det1} and \re{det3}. The analysis shows that, while the semiclassical expansion accurately describes the function $\Xi(z,k)$, it fails for $\mathcal{W}_n(z,k)$ when $n \ge 1$. 
 
\subsubsection*{Partition function} 
 
By combining the semiclassical expansion with numerical analysis and imposing certain nontrivial consistency conditions,  the grand canonical partition function $\Xi(z,k)$ defined in \re{Xi} and \re{det1} was conjectured
to have the following form at large $z = e^{\mu}$ and finite $k$, see \cite{Hatsuda:2015gca} for a review,
\begin{align}\label{Xi-conj}
\Xi(z,k) = \sum_{m=-\infty}^\infty e^{J(\mu + 2\pi i m, k)} \,,
\end{align}
where the grand potential $ J(\mu, k) $ decomposes into a perturbative and a nonperturbative part,
\begin{align}\label{J-grand}
J(\mu, k) = J^{\text{pert}}(\mu, k) + J^{\text{np}}(\mu, k) \,.
\end{align}
In contrast to the grand potential $J(\mu,k)$, the grand canonical partition function $\Xi(z,k)$ is a single-valued function of $z = e^{\mu}$ and must therefore remain invariant under the shift $\mu \to \mu + 2\pi i$.  
The summation in \re{Xi-conj} is required to restore periodicity and ensure the single-valuedness of $\Xi(z,k)$.

The perturbative contribution to \re{J-grand} is a cubic polynomial in $\mu = \log z$ with $k$-dependent coefficients,
\begin{align}\label{J-pt}
J^{\text{pert}}(\mu, k) = \frac{2}{3\pi^2 k}\, \mu^3 
+ \left( \frac{1}{3k} + \frac{k}{24} \right) \mu + A(k) \,,
\end{align}
where the constant term $A(k)$ admits an explicit integral representation \cite{Marino:2011eh,Hanada:2012si}.

The nonperturbative part is given by an infinite series of terms that are exponentially suppressed at large $\mu$
\begin{align}\label{Jnp}
J^{\text{np}}(\mu, k) = 
\sum_{\substack{m,\ell=0 \\ (m,\ell)\neq (0,0)}}^{\infty}
f_{m,\ell}(\mu)\, 
e^{-\left(\frac{4m}{k} + 2\ell\right)\mu} \,,
\end{align}
where the coefficients $ f_{m,\ell}(\mu) $ are polynomials in $\mu$.  

In the dual holographic description, the terms in \re{Jnp} with $\ell = 0$ and $m = 0$ correspond to contributions from worldsheet and membrane instantons, respectively, while the remaining terms in \re{Jnp} originate from bound states of these two types of instantons. The terms with $m \neq 0$ are genuinely nonperturbative and cannot be captured by the semiclassical expansion around $k = 0$. They were derived by exploiting the relation between the instanton coefficients $f_{m,0}$ and Gopakumar-Vafa invariants of the topological string. The contribution of the bound states in \re{Jnp} is not independent and it can be absorbed into the worldsheet instanton sector by redefining the chemical potential $\mu \mapsto \mu_{\rm eff}$.

Furthermore, the grand canonical partition function \re{Xi-conj} simplifies significantly for $k=1,2,4,8$ and the corresponding function $\Xi(N,k)$ can be written in a closed form in terms of the Jacobi theta functions \cite{Codesido:2014oua,Grassi:2014uua,Okuyama:2016xke}. 
  
Combining together the relations \re{Z,W} and \re{Xi-conj}, we obtain the following representation for the free energy of the ABJM theory at large $\mu$ and fixed $k$
\begin{align}\label{pre-Airy}
Z(N,k) = \int_{-i\infty}^{i\infty} {d\mu\over 2\pi i} e^{J(\mu,k)-N\mu}  \,.
\end{align}
As compared with \re{Z,W} the integration contour in this relation extends to infinity. Replacing the grand potential in \re{pre-Airy} with its expression \re{J-grand} and performing integration, each term of the instanton expansion of the partition function $Z(N,k)$ can be expressed in terms of Airy functions.

\subsubsection*{$\bm{1/6}$ BPS Wilson loop}

The expectation value of the $1/6$ BPS Wilson loop admits an integral representation analogous to \re{pre-Airy}
\begin{align}\label{W-pre-Airy}
W_n(N,k) = \int_{-i\infty}^{i\infty} \frac{d\mu}{2\pi i}\, \mathsf W_n(\mu,k)\, e^{J(\mu,k) - N\mu} \,.
\end{align}
The main difference between this relation and \re{Z,W} lies in the integration region. 
To reconcile the two, it suffices to decompose the integration contour in \re{W-pre-Airy} into a union of intervals $[i(2m-1)\pi,\, i(2m+1)\pi]$, with $m\in \mathbb Z$, and shift $\mu \to \mu + 2\pi m$ on each interval. 
Carrying out this procedure, we find that the function $\mathsf W_n(\mu,k)$ appearing on the right-hand side of \re{W-pre-Airy} 
is related to the generating function $\mathcal{W}_n(z,k)$ defined in \re{Xi} and \re{det3} by
\begin{align}\label{W-conj}
\mathcal{W}_n(z,k) = \frac{1}{\Xi(z,k)} \sum_{m=-\infty}^\infty \mathsf W_n(\mu + 2\pi i m,k)\, e^{J(\mu + 2\pi i m, k)} \,.
\end{align}
As in the case of $\Xi(z,k)$, the sum in \re{W-conj} restores the invariance of $\mathcal{W}_n(z,k)$ under 
$\mu \to \mu + 2\pi i$ and guarantees that the function is single-valued.

Previous studies have shown that, at large $\mu$ and finite $k$, the function $\mathsf W_n(\mu,k)$ admits the following general representation \cite{Klemm:2012ii,Hatsuda:2013yua,Okuyama:2016deu}
\begin{align}\label{W-dec}
\mathsf W_n(\mu,k) = \frac{i^n\, e^{2n\mu/k}}{k\, \sin\!\left(\frac{2\pi n}{k}\right)}
\left[\mathsf W^{\text{pert}}_n(\mu,k) +\mathsf W^{\text{np}}_n(\mu,k) \right],
\end{align}
where the prefactor is introduced for convenience.  
This normalization factor grows exponentially at large $\mu$ and develops a pole at $k = 2n$.  
The two terms inside the brackets in \re{W-dec} represent, respectively, the perturbative and nonperturbative contributions. Their structure closely parallels that of the grand potential \re{J-grand}. 

The perturbative part of \re{W-dec} is linear in $\mu$ 
\begin{align}\label{W-pt}
\mathsf W^{\text{pert}}_n(\mu,k)={\mu\over\pi} + C_n(k) -{ik\over 4}\,,
\end{align}
where $C_n(k)$ is a real constant. As compared with \re{J-pt}, this function develops an imaginary part.  The nonperturbative part is given by the sum over worldsheet and membrane instantons
\begin{align}\label{W-np}
\mathsf W_n^{\text{np}}(\mu, k) = 
\sum_{\substack{m,\ell=0 \\ (m,\ell)\neq (0,0)}}^{\infty}
w_{m,\ell}(\mu)\, 
e^{-\frac{4m}{k} \mu- 2\ell\mu} \,,
\end{align}
where the complex valued coefficients $w_{m,\ell}(\mu)$ depend on the winding number $n$ and are linear in $\mu$.

Thus, the problem of computing the $1/6$ BPS Wilson loop reduces to determining the constant term $C_n(k)$ in the perturbative contribution \re{W-pt} and the instanton coefficients $w_{m,\ell}(\mu)$ in the nonperturbative contribution \re{W-np}.

The perturbative contribution \re{W-pt} has been computed using two complementary approaches: through the semiclassical expansion of the Fermi gas ensemble \cite{Klemm:2012ii} and via high-precision numerical evaluation of the winding Wilson loop \cite{Okuyama:2016deu}.  
The corresponding constant term was found to be different  
\begin{align}\notag\label{A-mismatch}
{}& C^{\text{WKB}}_n(k) = -\cot\!\left(\frac{2\pi n}{k}\right) - \frac{k}{2\pi}\, H_{n-1}\,, \\
{}& C^{\text{num}}_n(k) = -\sum_{j=1}^n \cot\!\left(\frac{2\pi j}{k}\right),
\end{align}
where $H_n$ denotes the harmonic number. The two expressions agree for $n = 1$ but differ for $n \ge 2$.

This discrepancy reflects a general limitation of the semiclassical approach discussed above. Recall that the localization integral \re{Z}, and consequently the generating function $\mathcal{W}_n(z,k)$ defined in \re{Xi}, are well-defined only for $k > 2n$, developing a singularity at $k = 2n$. Therefore, for $n \ge 1$, the function $\mathcal{W}_n(z,k)$ cannot be analytically continued to the small-$k$ region without encountering a singularity. As explained above, this obstruction invalidates the applicability of the semiclassical approximation in this case. Note, however, that this argument does not apply for $n = 0$, where $\mathcal{W}_0(z,k)$ is related to the generating function $\Xi(z,k)$ through \re{W_0}.  
This explains why the semiclassical approximation remains valid for the partition function, but fails for the Wilson loop.

The imaginary and real parts of the nonperturbative contribution to the $1/6$ BPS Wilson loop \re{W-np} were studied numerically in \cite{Hatsuda:2013yua,Okuyama:2016deu} for the first few winding numbers, $ n = 1, 2, 3, 4 $.~\footnote{In virtue of \re{W-1/2},  the function $ \Im \mathsf W_n^{\text{np}}(\mu, k) $ coincides with the nonperturbative part of the $1/2$ BPS Wilson loop. Because of the normalization factor in \re{W-dec}, the function $ \Re \mathsf W_n^{\text{np}}(\mu, k) $ corresponds to the imaginary part of the $1/6$ BPS Wilson loop in the conventions of \cite{Okuyama:2016deu}.} By evaluating $\mathsf W_n^{\text{np}}(\mu, k) $ numerically for some positive integer values of $ k $, the exact values of the instanton coefficients $w_{m,\ell}(\mu) $ were guessed for the lowest values of $ m $ and $ \ell $ through numerical fitting. Building on this analysis, a number of conjectures were formulated concerning the properties of the nonperturbative function $  \mathsf W_n^{\text{np}}(\mu, k) $.

It is worth emphasizing that, despite extensive numerical investigations, the large-$\mu$ expansion of the generating functions \re{Xi-conj} and \re{W-conj} -- or, equivalently, the large-$N$ expansion of the partition function and the winding Wilson loop -- remains a conjecture. At present, no analytical method is known for deriving these functions directly from their operator representations \re{det1} and \re{det3}. The purpose of this paper is to develop such a method.
 
\section{Tracy-Widom approach}\label{sect3}
 
In this section, we present an alternative approach to computing the generating functions \re{det1} and \re{det3}. Its main advantage over the semiclassical expansion is that it does not rely on any approximation, and it provides a powerful exact representation of these functions that remains valid for arbitrary values of the parameters $N$, $k$, and $n$. 

This approach was originally developed by Tracy and Widom \cite{Tracy:1995ax}, who proved the conjecture formulated in \cite{Cecotti:1992qh,Zamolodchikov:1994uw}, according to which the Fredholm determinant of an integral operator of the same type as in \re{det1} satisfies a TBA-like system of nonlinear equations.
The integral operator in \re{det1} involves the kernel $\rho(x,y)$ defined in \re{rho-fun}. It  represents a special case of the more general class of kernels analyzed in~\cite{Tracy:1995ax}.~\footnote{Note that the kernel $\rho(x,y)$ is not symmetric, $\rho(x,y)\neq\rho(y,x)$. To bring it to the form considered in~\cite{Tracy:1995ax}, it has to be symmetrized by a similarity transformation, which leaves the Fredholm determinant~\re{det1} invariant.}

We begin by introducing an auxiliary function which plays a central role in our analysis~\footnote{In the notation of \cite{Tracy:1995ax}, this function is given by the sum of functions $Q(\theta)$ and $P(\theta)$ for $\theta=x/k$.  }
\begin{align}\label{psi}
\psi(x|z) =\VEV{E\Big|{1\over 1-z \bm\rho}\Big|x} \,,  
\end{align}
where the operator $\bm\rho$ is defined in \re{rho-oper} and the notation was introduced for $E(x)=\sqrt{2}\, e^{x\over 2k}$. At small $z$ the function \re{psi} admits an expansion
\begin{align}\notag\label{psi-small-z}
{}& \psi(x|z)=\sum_{\ell\ge 0} z^\ell \psi_\ell(x)\,,
\\
{}&\psi_{\ell+1}(x) = \int_{-\infty}^\infty dy\, \psi_\ell(y)  \rho(y,x)\,,
\end{align}
where $\psi_0(x)=\sqrt 2\, e^{x\over 2k}$ and the kernel $\rho(x,y)$ is given by \re{rho-fun}. 
 
The rationale for introducing the auxiliary function \re{psi} is that the generating function \re{det3} can be expressed in terms of this function 
\begin{align}\notag\label{W-psi}
\mathcal W_n(z,k) {}&= {i z\over 2\sin({2\pi n\over k}) } {e^{i\pi n(n-1)/k}\over 4\pi k}\int_{-\infty}^\infty dx \, e^{(n-1)x\over k}\chi(x)
\\[2mm]
{}& \times \left[e^{-{i\pi n\over k}}\psi(x|z)\psi(x+2i\pi n|-z) -e^{{i\pi n\over k}} \psi(x|-z)\psi(x+2i\pi n|z)\right],
\end{align}
where $\chi(x)=1/(2\cosh(x/2))$.
For $n\to 0$ this relation can be combined with \re{W_0} to obtain the derivative of the generating function $\Xi(z,k)$. 
The derivation of the relation \re{W-psi} can be found in Appendix~\ref{app:A}. 

Replacing the $\psi-$function in \re{W-psi} with its small $z$ expansion \re{psi-small-z}, we can systematically expand 
the generating function $\mathcal W_n(z,k)$ in powers of $z$ and reproduce the relation \re{W-small-z}. To find  $\mathcal W_n(z,k)$ for arbitrary $z$, we have to determine the function \re{psi}. This will be done in two steps. First, we formulate a set of nontrivial relations the function $\psi(x|z)$ has to satisfy and, then, present the solution.

We emphasize that the relation~\re{W-psi} is exact and holds for arbitrary values of the parameters.
Note the presence of the sine factor in the denominator of \re{W-psi}, which gives rise to a singularity at $k = 2n$.
This behavior is consistent with the earlier observation \re{k>2n} that the localization integral  \re{Z} is well defined only for $k > 2n$.

\subsection{Properties of $\psi-$function}

The function~\re{psi} can be expanded over the eigenspectrum of the operator~$\bm\rho$
\begin{align}\label{rho-Sch}
\int_{-\infty}^\infty dy\, \rho(y,x) \phi_s(y) = \lambda_s \phi_s(x)\,.
\end{align}
This operator possesses a discrete spectrum of real, positive eigenvalues that accumulate at the origin \cite{Hatsuda:2012hm}.
These eigenvalues can be conveniently parameterized as $\lambda_s=e^{-\varepsilon_s}$, where $\varepsilon_s$ are monotonically increasing with $s$. 

Using the operator representation \re{rho-oper}, one finds that the eigenfunctions satisfy a finite-difference equation (see \cite{Marino:2016new})
\footnote{A conjecture for the exact eigenfunctions, based on the topological string/spectral theory correspondence, was proposed in \cite{Marino:2016rsq}.}
\begin{align}\label{phi-eig}
\phi_s(x+i\pi k) +\phi_s(x-i\pi k) = e^{\varepsilon_s}\chi(x)\phi_s(x)\,,
\end{align}
where the function $\chi(x)$ was defined in \re{W-psi}. 
The $\psi-$function \re{psi} can be expanded over these eigenfunctions as
\begin{align}\label{psi-sum}
\psi(x|z) = \sum_{s\ge 0} c_s {\phi_s(x) \over 1-z e^{-\varepsilon_s}}\,,
\end{align}
with the expansion coefficients $c_s$ independent of $x$ and $z$. It then follows from the last relation that $\psi(x|z)$ is a meromorphic function of~$z$ with poles located at the real positive $z = e^{\varepsilon_s}$. 
 
Similarly, the Fredholm determinant \re{det1} can be written as  
\begin{align}
\Xi(z,k) = \prod_{s \ge 0} \bigl(1 + z e^{-\varepsilon_s}\bigr)\,.
\end{align}
This function is entire in $z$ and has zeros located on the negative real axis, at the positions of the poles of the function $\psi(x|-z)$.  
For reasons that will become clear shortly, it is convenient to introduce the ratio of determinants  
\begin{align}\label{D-def}
D(z) = \frac{\Xi(z,k)}{\Xi(-z,k)} = \prod_{s \ge 0} \frac{1 + z e^{-\varepsilon_s}}{1 - z e^{-\varepsilon_s}}\,.
\end{align}
Its dependence on $k$ is understood implicitly. 

The function $D(z)$ possesses an infinite set of poles and satisfies the identity $D(z)\,D(-z)=1$.
Consequently, its logarithm is an odd function of $z$,  
\begin{align}\label{F-def}
 \mathcal F(z)=-\mathcal F(-z) = \log D(z) \,.
\end{align}
It exhibits an infinite sequence of logarithmic branch cuts in the complex $z-$plane, starting at $z = \pm e^{\varepsilon_s}$ for $s \ge 0$. With a slight abuse of notation, we will refer to the function $\mathcal F(z)$ as the parity-odd free energy.

We can use the relations \re{D-def} and \re{W_0} to derive the following identity for the function~$\mathcal{F}(z)$
\begin{align}\label{dF-W}
z \, \partial_z \mathcal{F}(z) 
= z \, \partial_z \log \frac{\Xi(z,k)}{\Xi(-z,k)} 
= \mathcal{W}_0(z,k) - \mathcal{W}_0(-z,k)\,.
\end{align}
Substituting the representation \re{W-psi} of $\mathcal{W}_0(z,k)$, we find
\begin{align}\label{dF}
z \, \partial_z \mathcal{F}(z) = \frac{z}{4\pi k} \int_{-\infty}^\infty dx \; \chi(x)\, e^{-x/k} \, \psi(x|-z)\, \psi(x|z)\,.
\end{align}
This relation has a structure similar to~\re{W-psi}. It provides an explicit equation for the ratio of determinants~\re{D-def} in terms of the $\psi-$function.

\subsection{Baxter equation}

In the previous subsection, we analyzed the analytical properties of the function $\psi(x|z)$ in the complex $z-$plane. We now turn to its properties as a function of $x$.

For this purpose it is convenient to use the representation \re{psi-small-z} and examine the recurrence relation between the functions $\psi_\ell(x)$
\begin{align}\label{psi-ell} 
\psi_{\ell+1}(x) = {1\over 8\pi k}\int_{-\infty}^\infty {dy\, \psi_\ell(y)\over\cosh({x-y\over 2k})\cosh(\frac{y}2)}\,,
\end{align}
where we replaced the kernel $\rho(y,x)$ with its expression \re{rho-fun}. Note that the right-hand side acquires a sign factor $(-1)^m$ under the shift $x\to x+2\pi im k$ with $m$ being integer. The same property holds for the function $\psi_0(x) =\sqrt 2 e^{x\over 2k}$. Together with \re{psi-small-z} this leads to the periodicity condition
\begin{align}\label{per}
\psi(x+2\pi im k|z)=(-1)^m \psi(x|z)\,,  
\end{align}
where $m \in \mathbb Z$. This relation holds for arbitrary complex $x$.

For complex $x$, the integral on the right-hand side of \re{psi-ell} is well-defined for $-\pi k < \Im x < \pi k$ and, as a consequence, the function $\psi_{\ell+1}(x)$ is analytical within this strip. At the same time, for $\Im x \to \pm\pi k$ the integral in \re{psi-ell} develops a logarithmic singularity from integration in the vicinity of $y=\Re x$. Together with \re{per} this implies that the function $\psi(x|z)$ has an infinite number of logarithmic branch cuts at $\Im x=(2m-1)\pi k$ with $m \in \mathbb Z$. 
These cuts run parallel to the real axis and are separated by $2\pi k$ along the imaginary axis. Notice that 
in the semiclassical limit $k \to 0$, discussed in the previous section, the analyticity strip $-\pi k < \Im x < \pi k$ shrinks to the real axis.

We can use \re{psi-ell} to evaluate the discontinuity of $\psi_{\ell+1}(x)$ across the cut at $\Im x=\pi k$. For real $x$ and $\epsilon \to 0$ we have 
\begin{align}\notag\label{pre-Bax}
{}&  \psi_{\ell+1}(x+ i\pi (k-\epsilon))-\psi_{\ell+1}(x+ i\pi (k+\epsilon)) 
 \\[2mm]
{}& \qquad =   \psi_{\ell+1}(x+ i\pi (k-\epsilon))+\psi_{\ell+1}(x- i\pi (k-\epsilon))  = \chi(x) \psi_\ell(x) \,,
\end{align}
 where in the first relation we applied  \re{per}. It then follows from \re{psi-small-z} that the function $\psi(x|z)$ satisfies the Baxter equation
\begin{align}\label{Bax}
\psi(x+i\pi k|z) + \psi(x-i\pi k|z) = z \chi(x) \psi(x|z) \,.
\end{align}
This relation holds for real $x$. Note that the relation \re{Bax} closely resembles \re{phi-eig}. Indeed, by substituting the expansion \re{psi-sum} into the Baxter equation~\re{Bax}, one readily recovers \re{phi-eig}. 

It is important to emphasize that the left-hand side of \re{Bax} is interpreted in the same manner as in the second line of~\re{pre-Bax}. Namely, the functions $\psi(x+i\pi k|z)$ and $\psi(x-i\pi k|z)$ are defined just below and just above the branch cuts, respectively (see Figure~\ref{fig:poles} below).
 
\subsection{Wronskian relation}

It follows from the definition \re{psi} that the function $\psi(x|z)$ obeys the differential equation
\begin{align}\label{dpsi}
z\partial_z \psi(x|z) = \VEV{E\Big|{z \bm\rho\over (1-z \bm\rho)^2}\Big|x}=-{1\over 4\pi k} \int_{-\infty}^\infty dy\,\chi(y)\psi(y|z) \Gamma(y,x|-z)\,,
\end{align}
where the notation was introduced for the resolvent kernel of the operator~\re{rho-oper},
\begin{align}\label{Gamma}
\Gamma(x,y|z) = \frac{4\pi k}{\chi(x)} \VEV{x\left| \frac{z\, \boldsymbol{\rho}}{1 + z\, \boldsymbol{\rho}} \right| y}.
\end{align}
The prefactor is chosen to ensure that the function is symmetric under the exchange of its arguments (see \re{R-psi} below).

Using the definition of the operator~$\boldsymbol{\rho}$ in~\re{rho-oper}, we obtain an equivalent representation of \re{Gamma}
\begin{align}\label{Gamma-K}
\Gamma(x,y|z) = z \VEV{x\left| \boldsymbol{K} \frac{1}{1 + z\, \boldsymbol{\rho}} \right| y},
\end{align}
where the operator $\boldsymbol{K} = 2\pi k/\cosh(\boldsymbol{p}/2)$ has the kernel
\begin{align}\label{K-ker}
K(x,y)=\vev{x|\boldsymbol{K}|y} = \frac{1}{\cosh \left(\frac{x-y}{2k}\right)}\,.
\end{align}
Note that this kernel develops a pole at $x - y = i\pi k$, a feature that will play an important role below.

Using operator identities for~$\boldsymbol{\rho}$, we can show (see Appendix~\ref{app:A}) that the function~\re{Gamma} admits the representation  
\begin{align}\label{R-psi}
\Gamma(x,y|z) = \frac{z}{2\sinh\left(\frac{x-y}{k}\right)}
\Bigl[e^{-y/k}\psi(x|z)\psi(y|-z) - e^{-x/k}\psi(x|-z)\psi(y|z)\Bigr].
\end{align}
This function is symmetric, $\Gamma(x,y|z)=\Gamma(y,x|z)$. Moreover, it follows from \re{Gamma-K} that it is meromorphic in~$z$, with poles at $z=-e^{\varepsilon_s}$ (see~\re{rho-Sch}).  
Since the function $\psi(x|z)$ has poles at $z=e^{\varepsilon_s}$, the residues of $\Gamma(x,y|z)$ at these poles must vanish for \re{R-psi} to hold.  

Because of the prefactor in~\re{R-psi}, the function $\Gamma(x,y|z)$ develops a pole at $x-y=i\pi k$.  
The same singularity appears in the kernel~\re{K-ker} of the operator~$\boldsymbol{K}$, as expected from the small-$z$ expansion of~\re{Gamma-K},
\begin{align}\label{Gamma-exp}
\Gamma(x,y|z) = z\VEV{x|\boldsymbol{K}|y} - z^2 \VEV{x|\boldsymbol{K}\boldsymbol{\rho}|y} + O(z^3)\,.
\end{align}
Inspection of the integral representation of the matrix elements on the right-hand side shows that the pole at $x-y=i\pi k$ arises solely from the $O(z)$ term.  
Matching the residues of~\re{R-psi} and~\re{Gamma-exp} at this pole then yields the relation  
\begin{align}\label{Wr}
\psi\!\left(x+\tfrac{i\pi k}{2}\Big|z\right)\psi\!\left(x-\tfrac{i\pi k}{2}\Big|-z\right)
+\psi\!\left(x+\tfrac{i\pi k}{2}\Big|-z\right)\psi\!\left(x-\tfrac{i\pi k}{2}\Big|z\right)
= 4\,e^{x/k}\,.
\end{align}
It can be viewed as a Wronskian-type relation for the solutions of the Baxter equation~\re{Bax}.  
Unlike~\re{Bax}, however, equation~\re{Wr} holds for arbitrary complex~$x$.  
As in~\re{R-psi}, both sides of \re{Wr} must have vanishing residues at the poles of the $\psi$-functions located at $z=\pm e^{\varepsilon_s}$.

\subsection{Shift relation}

Equation~\re{Wr} relates the $\psi-$functions whose arguments differ by a shift of $i\pi k$.  
In this subsection, we derive a stronger relation that expresses $\psi(x+2\pi i|z)$ as a linear combination of $\psi(x|z)$ and $\psi(x|-z)$.  

To this end, we replace $x \to x + 2\pi i m$ (with $m \in \mathbb{Z}$) on both sides of \re{psi-ell} and simultaneously shift the integration variable as $y \to y + 2\pi i m$. This yields  
\begin{align}\label{psi-ell1} 
\psi_{\ell+1}(x+2\pi i m) 
= \frac{(-1)^m}{8\pi k} 
\int_{-\infty - 2\pi i m}^{\infty - 2\pi i m} 
\frac{dy\, \psi_\ell(y + 2\pi i m)}{
\cosh\!\left(\frac{x - y}{2k}\right)
\cosh\!\left(\frac{y}{2}\right)}\,,
\end{align}
where the integration contour is shifted into the complex plane. In the next step, we deform the contour back to the real axis and collect the residues of the integrand enclosed within the strip $-2\pi m < \Im y < 0$.

\begin{figure} 
\tikzset{
  arrowcircle/.style={
    red, line width=1.2pt,
    postaction={decorate},
    decoration={markings, mark=at position 0.95 with {\arrow{>}}}
  },
  cross/.style={line width=1.2pt, blue},
  axis/.style={line width=0.5pt, ->, >=Stealth},
  dashedline/.style={black, dashed,line width=1.2pt}
}

\begin{center} 
\begin{tikzpicture}[x=0.5cm,y=0.5cm,scale=0.8]
  % Axes
  \draw[axis] (-11,0) -- (11,0); % horizontal
  \draw[axis] (0,-10) -- (0,10); % vertical
  % Dashed horizontal lines at y = ±9.5
  \draw[dashedline] (-11,9.5) -- (11,9.5);
  \draw[dashedline] (-11,-9.5) -- (11,-9.5);
  % Red thick horizontal arrow at y = -8.5
  \draw[red, very thick, -{Stealth}] (-11,-8.25) -- (11,-8.25);
 % Red thick horizontal arrow at y = 0
  \draw[blue, very thick, -{Stealth}] (-11,0) -- (11,0);
  % Cross symbol
  \newcommand{\drawcross}[1]{%
    \draw[cross] (-0.2,#1-0.2) -- (0.2,#1+0.2);
    \draw[cross] (-0.2,#1+0.2) -- (0.2,#1-0.2);}
  % Crosses and labels
  \drawcross{1.5}
  \node[right, xshift=4pt] at (0,1.5) {${}$};
  \drawcross{3.75}
  \node[right, xshift=4pt] at (0,3.75) {${}$};
  \drawcross{6}
  \node[right, xshift=4pt] at (0,6) {${}$};
  \drawcross{-1.5}
  \node[right, xshift=4pt] at (0,-1.5) {${}$};
  \draw[arrowcircle] (0,-1.5) circle (0.65);
  \drawcross{-3.75}
  \node[right, xshift=4pt] at (0,-3.75) {${}$};
  \draw[arrowcircle] (0,-3.75) circle (0.65);
  \drawcross{-6}
  \node[right, xshift=4pt] at (0,-6) {${}$};
  \draw[arrowcircle] (0,-6) circle (0.65);
  % Markers at horizontal-line crossings with labels
 % \fill (0,9.5) circle (0.8pt);
  \node[right, xshift=4pt] at (0,9) {$\pi k$};
 % \fill (0,-9.5) circle (1.8pt);
  \node[right, xshift=2pt] at (0,-8.85) {$-\pi k$};
 % \fill (0,-8.5) circle (0.8pt);
  \node[right, xshift=2pt] at (0,-7.5) {$-2\pi m$};
\node[right, xshift=2pt] at (0,0.75) {$0$};
\end{tikzpicture}
\end{center}
\caption{The function $\psi(x|z)$ is analytic in the complex $x$-plane within the strip $-\pi k < \Im x < \pi k$, whose boundaries are indicated by dashed lines. The blue and red lines denote the integration contours appearing in relations~\re{psi-ell} and~\re{psi-ell1}, respectively. In going from relation~\re{psi-ell1} to~\re{sh-1}, the red contour is deformed so as to coincide with the blue one, thereby enclosing the poles indicated by blue crosses, whose residues are then picked up.
}\label{fig:poles}
\end{figure}
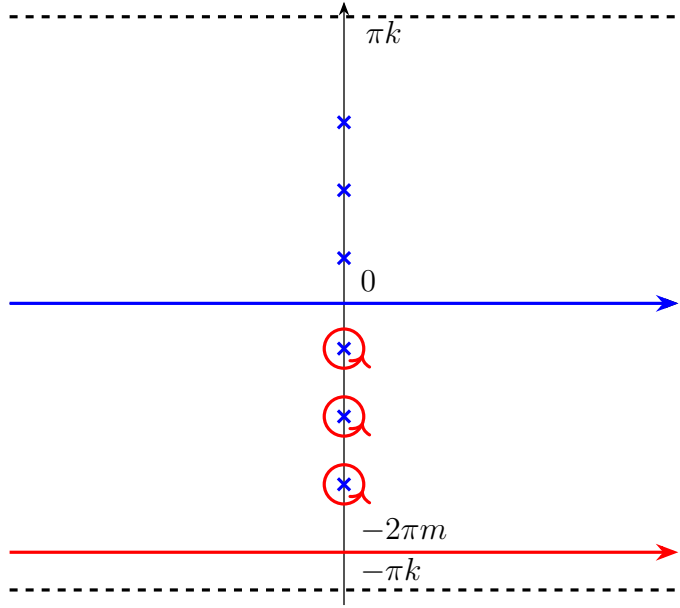

We have seen previously that the function $\psi_\ell(x)$ is analytic in the strip 
$-\pi k < \Im x < \pi k$. 
Therefore, for $0 < m < k/2$, the function $\psi_\ell(y + 2\pi i m)$ has no poles within the strip 
$-2\pi m < \Im y < 0$. 
The same is true for the first $\cosh$ function in the denominator of~\re{psi-ell1}. 
Hence, the only poles of the integrand originate from the zeros of $\cosh(y/2)$ in the denominator of \re{psi-ell1}, 
which are located at $y = -i\pi(2s + 1)$ for $0 \le s \le m - 1$. 
Evaluating the residues at these poles, we obtain from \re{psi-ell1}
\begin{align}\notag\label{sh-1}
\psi_{\ell+1}(x + 2\pi i m)
&= (-1)^m \int_{-\infty}^{\infty} dy\, \rho(y, x)\, \psi_\ell(y + 2\pi i m)
\\
&  - \frac{1}{2k} \sum_{s=0}^{m-1} (-1)^{m+s} 
K(-i\pi(2s + 1), x)\, \psi_\ell(i\pi(2(m - s) - 1))\,,
\end{align}
where the kernels $\rho(y, x)$ and $K(x, y)$ are defined in~\re{rho-fun} and~\re{K-ker}, respectively. This relation is illustrated in Figure~\ref{fig:poles}.

Multiplying both sides of~\re{sh-1} by $z^{\ell+1}$ and summing over $\ell \ge 0$, we find, using~\re{psi-small-z},
\begin{align}\notag\label{sh-2}
\psi(x + 2\pi i m | z) 
&- (-1)^m z \int_{-\infty}^{\infty} dy\, \rho(y, x)\, \psi(y + 2\pi i m | z)
\\
&= \psi_0(x + 2\pi i m)
- \frac{z}{2k} \sum_{s=0}^{m-1} (-1)^{m+s} 
K(-i\pi(2s + 1), x)\, \psi(i\pi(2(m - s) - 1) | z)\,,
\end{align}
where $\psi_0(x + 2\pi i m) = e^{i\pi m / k} E(x)$ is defined in~\re{psi-small-z}.

Note that the first line of~\re{sh-2} involves a convolution of the function $\psi(x + 2\pi i m | z)$ with the kernel of the operator $1 - (-1)^m z\,\bm{\rho}$.  
Applying the inverse operator to both sides of this relation and using~\re{psi} and~\re{Gamma-K}, we obtain
\begin{align}\notag\label{shift}
{}& \psi(x + 2\pi i m \,|\, z) 
= e^{\frac{i\pi m}{k}} \psi(x \,|\, (-1)^m z) 
\\
{}&\qquad 
+ \frac{1}{2k} \sum_{s=0}^{m-1} (-1)^{s} 
\Gamma\!\left(x, -i\pi(2s+1) \,\middle|\, (-1)^{m+1}z\right) 
\psi\!\left(i\pi [2(m - s) - 1] \,\middle|\, z\right).
\end{align}
This relation holds for $1 \le m < k/2$ and for arbitrary complex $x$.  
Applying complex conjugation to both sides of~\re{shift} yields the corresponding relation for $\psi(x - 2\pi i m, z)$.
  
Replacing the $\Gamma-$function by its representation~\re{R-psi}, we find that the right-hand side of~\re{shift} can be written as a linear combination of the functions $\psi(x|(-1)^m z)$ and $\psi(x|(-1)^{m+1} z)$, with coefficients involving special values of these function evaluated at $x = \pm i\pi(2s + 1)$ for $0 \le s \le m - 1$.  
In the particular case $m=1$, this expression simplifies to  
\begin{align}\notag\label{id2}
\psi(x + 2\pi i|z) 
&{} = z\, e^{\frac{i\pi}{k}}
   \frac{\psi(i\pi|z)\psi(-i\pi|-z)}{4k\, \sinh\!\left(\frac{x + i\pi}{k}\right)} \,
   \psi(x|z)  \\ 
&{}
+ \bigg[
   e^{\frac{i\pi}{k}}
   - z e^{-\frac{x}{k}}
     \frac{\psi(i\pi|z)\psi(-i\pi|z)}{4k\, \sinh\!\left(\frac{x + i\pi}{k}\right)}
  \bigg]
  \psi(x|-z)\,.
\end{align}
The coefficients of $\psi(x|z)$ and $\psi(x|-z)$ on the right-hand side depend on $\psi(i\pi|z)$, 
$\psi(-i\pi|z)$ and $\psi(-i\pi|-z)$. As we will see in the next section, these special values of the $\psi-$function play a special role in our analysis. 

\subsection{Parity relation} 

We observe that the relations \re{R-psi}, \re{Wr}, and \re{id2} involve the functions $\psi(x|z)$ and $\psi(x|-z)$.
These two functions have different analytic properties in the complex $z-$plane. Namely, $\psi(x|z)$ has poles on the positive real semi-axis, while $\psi(x|-z)$ has poles on the negative real semi-axis. It is therefore not obvious that these two functions are related.

We show in Appendix~\ref{app:B} that, for arbitrary complex values of $x$ and $z$, the functions $\psi(-x|z)$ and $\psi(x|-z)$ satisfy the relation
\begin{align}\label{par}
\psi(-x|z) = e^{-\frac{x}{k} + \mathcal F(z)} \, \psi(x|-z)\,,
\end{align}
where the free energy $\mathcal{F}(z)$ is defined in \re{F-def}. It follows from \re{D-def} that the exponential factor $e^{\mathcal{F}(z)}$ on the right-hand side of \re{par} precisely cancels the poles of $\psi(x|-z)$ and reproduces those of $\psi(-x|z)$, ensuring consistency of their analytic structure.

In what follows, we employ the relation~\re{par} to express $\psi(x|-z)$ in terms of $\psi(-x|z)$. For instance, the relation \re{R-psi} can be rewritten as 
\begin{align}\label{R-psi1}
\Gamma(x,y|z) = \frac{z\, e^{- \mathcal F(z)} }{2\sinh\left(\frac{x-y}{k}\right)}
\Bigl[\psi(x|z)\psi(-y|z) - \psi(-x|z)\psi(y|z)\Bigr].
\end{align}

\section{Quantization condition}\label{sect4}

In the previous section, we expressed the generating function~\re{W-psi} in terms of the auxiliary $\psi-$function and derived a set of nontrivial relations satisfied by this function. These relations are exact and hold for arbitrary values of the Chern--Simons  level~$k$, the winding number~$n$, and the fugacity parameter~$z$.

In this section, we employ these relations to determine the special values of $\psi(x|z)$ along the imaginary $x-$axis, at the points $x = i\pi(2m + 1)$ with $m \in \mathbb{Z}$.
The motivation for computing these values is that, as will be shown in the next section, they are required for the evaluation of the generating function~\re{W-psi}.

Let us begin with $\psi(\pm i\pi|z)$. Since $\psi(x|z)$ is a real function of $x$, we can parameterize it as
\begin{align}\label{rho}
\psi(i\pi|z) = r(z)\, e^{i\Phi(z)/2}\,, \qqqquad
\psi(-i\pi|z) =r(z)\, e^{-i\Phi(z)/2}\,,
\end{align}
where the dependence of $r(z)$ and $\Phi(z)$ on the parameter $k$ is implicit.
Substituting $x = i\pi$ into \re{par} and using \re{rho}, we find
\begin{align}\label{Phi-par}
r(-z) = e^{-\mathcal F(z)}\, r(z)\,, \qqqquad
\Phi(-z) = \frac{2\pi}{k} - \Phi(z)\,.
\end{align}
We show below that the functions $r(z)$ and $\Phi(z)$ can be expressed in terms of $\mathcal F(z)$ (see \re{rho-F} and \re{QCs}).

Having determined the functions $r(z)$ and $\Phi(z)$, we can use \re{id2} with $x = i\pi$ to express $\psi(3i\pi|z)$ and its complex conjugate in terms of the values in \re{rho}
\begin{align}\label{psi3}
\psi(3i\pi|z) =e^{\frac{i \pi }{k}} \psi (i \pi|-z)+i\frac{z\, \psi (i \pi|z)}{4 k \sin \left(\frac{2 \pi }{k}\right)} \left[{e^{-\frac{i \pi }{k}} \psi (-i \pi|z) \psi (i \pi|-z)-e^{\frac{ i \pi }{k}}\psi (-i \pi |-z) \psi (i
   \pi|z) }\right].
\end{align}
By iterating this procedure for $x = 3i\pi, 5i\pi, \dots$, we can recursively compute the values of $\psi(\pm i\pi(2m + 1)|z)$ for $m\ge 2$ using \re{rho} as the initial condition. These values will be essential in what follows.
 
\subsection{Consistency relations}

By combining the relations \re{id2} and \re{par}, the function $\psi(x+2i\pi|z)$ can be expressed as a linear
combination of $\psi(x|z)$ and $\psi(-x|z)$
\begin{align}\label{A,B}
\psi(x+2i\pi|z) = A(x)\psi(x|z) + B(x) \psi(-x|z)\,,
\end{align}
where the notation was introduced for functions
\begin{align}\notag\label{A+B} 
A(x){}&=\frac{z\, r^2(z)}{4 k \sinh(\frac{x+i \pi }{k})} e^{-\mathcal F(z)+i\Phi(z)} \,,
\\ 
B(x){}&=e^{-\mathcal F(z)} \left[ e^{\frac{x+i \pi }{k}}-\frac{z\, r^2(z)}{4k\sinh(\frac{x+i \pi }{k})} \right]. 
\end{align}
For simplicity, the dependence of these functions on $z$ is not shown explicitly. 

The relation~\re{A,B} imposes nontrivial constraints on the functions $A(x)$ and $B(x)$.
To make this explicit, we first replace $x \to -x$ in \re{A,B}, take the complex conjugate, and combine the resulting equation with the original one. The resulting system can then be recast in matrix form,
\begin{align}\label{2psi}
\lr{\psi(x+2i\pi|z)\atop \psi(-x-2i\pi|z)} = T(x) \lr{\psi(x|z)\atop \psi(-x|z)} \,, 
\end{align}
where the matrix $T(x)$ encodes the action of a shift $x\to x+2i\pi$ on the two-component vector of $\psi-$functions. Its matrix elements depend on the functions $A(x)$ and $B(x)$ and their complex conjugates
\begin{align}\label{U}
T(x) =\left[  \begin{array}{cc}A(x)  & B(x) \\ \bar B(-x) & \bar A(-x) \end{array} \right].
\end{align}
 
Next, we replace $x \to x - 2i\pi$ in \re{2psi} and take the complex conjugate to obtain another relation among the same $\psi-$functions. Requiring consistency between the two matrix equations leads to the conditions
\begin{align}\notag\label{AB}
{}& A(x-2i\pi) \bar B(x) + B(x-2i\pi) A(-x)=0 \,, 
\\[1.5mm]
{}&  A(x-2i\pi) \bar A(x) + B(x-2i\pi) B(-x)=1\,.
\end{align}
Substituting the expressions for $A(x)$ and $B(x)$ from \re{A,B}, we find that the first relation is satisfied automatically, while the second yields
\begin{align}\label{rho-F}
r^2(z)={2k\over z} \lr{e^{2\mathcal F(z)}-1}\,.
\end{align}
Thus, the absolute value of $\psi(\pm i\pi|z)$ is fixed by the function $\mathcal F(z)$ introduced in \re{F-def}. 

In the next step, we need to determine the phase $\Phi(z)$ defined in~\re{rho}. The subsequent analysis then proceeds differently for odd and even values of $k$.

\subsection{Phase at even $k$}  

Let us start with the Baxter equation \re{Bax} and put $x=0$
\begin{align}\label{Bax1}
 \psi(i\pi k|z)+\psi(-i\pi k|z)= \frac12 z \psi(0|z)\,.
\end{align} 
For even $k$, we can apply the shift relation \re{2psi} recursively $k/2$ times to get
\begin{align}
\lr{\psi(i\pi k|z)\atop \psi(-i\pi k|z)} = \mathbb T_{\rm e}\lr{\psi(0|z)\atop \psi(0|z)} \,,
\end{align} 
where the notation was introduced for the transfer matrix
\begin{align}\label{Te}
\mathbb T_{\rm e} = T(i\pi(k-2)) \dots T(2i\pi)T(0)
\end{align}
and subscript refers to even value of $k$.

It follows from the Baxter equation \re{Bax1} that the transfer matrix has to satisfy the following quantization condition
\begin{align}\label{QC-e}
(1,1)\, \mathbb T_{\rm e}  \lr{1 \atop 1} 
= \frac12 z\,.
\end{align}
Being combined with \re{U} and \re{A+B} it leads to the equation for the phase $\Phi(z)$. The solution to this equation is discussed below.
 
\subsection{Phase at odd $k$} 

Substituting $x = -i\pi k/2$ into the Wronskian relation~\re{Wr} and using~\re{par}, we find  
\begin{align}
\psi(0|z)\,\psi(i\pi k|z) - \psi(0|z)\,\psi(-i\pi k|z)
= 4i\, e^{\mathcal F(z)}\,,
\end{align}
where we  used $\chi(0) = 1/2$.  
Combining this relation with the Baxter equation~\re{Bax1} and eliminating $\psi(0|z)$, we obtain
\begin{align}\label{qc-o}
\psi^2(i\pi k|z) - \psi^2(-i\pi k|z)
= 2i z\, e^{\mathcal F(z)}\,.
\end{align}

As in the even-$k$ case, the relation~\re{2psi} can be employed to express the $\psi-$values on the left-hand side in terms of $\psi(\pm i\pi|z)$
\begin{align}\label{shift-o}
\begin{pmatrix}
\psi(i\pi k|z) \\[2mm] \psi(-i\pi k|z)
\end{pmatrix}
= \mathbb{T}_{\rm o}
\begin{pmatrix}
\psi(i\pi|z) \\[2mm] \psi(-i\pi|z)
\end{pmatrix}.
\end{align}
The corresponding transfer matrix reduces to  $\mathbb{T}_{\rm o} =1$ for $k=1$, while for $k\ge 3$ it is given by
\begin{align}\label{To}
\mathbb{T}_{\rm o} = T(i\pi(k-2)) \dots T(3i\pi)\,T(i\pi)\,.
\end{align}
In contrast to~\re{Te}, the argument of the $U-$matrices here involves only odd multiples of~$i\pi$.

Combining together \re{qc-o} and \re{shift-o}, we obtain the quantization condition at odd $k$
\begin{align}
(\psi(i\pi|z) \,,  \psi(-i\pi|z))(\mathbb{T}_{\rm o})^{\rm t} \sigma_3 \mathbb{T}_{\rm o}
\begin{pmatrix}
\psi(i\pi|z) \\[2mm] \psi(-i\pi|z)
\end{pmatrix}=2i z\, e^{\mathcal F(z)}\,,
\end{align}
where $\sigma_3$ is Pauli matrix and the superscipt `${\rm t}$' denotes the transposition.
We can simplify this relation further by applying \re{rho} and \re{rho-F}
\begin{align}\label{QC-o}
 \lr{e^{i\Phi/2},\ e^{-i\Phi/2}} \,(\mathbb{T}_{\rm o})^{\rm t} \sigma_3 \,\mathbb{T}_{\rm o} \,\lr{e^{i\Phi/2}\atop e^{-i\Phi/2}}= {i z^2 \over 2k\sinh \mathcal F(z)} \,.
\end{align}
This relation allows us to compute the phase $\Phi(z)$ in terms of the function $\mathcal F(z)$.

\subsection{Solutions to the quantization conditions} 

Let us now examine the quantization conditions~\re{QC-e} and~\re{QC-o} for several specific values of~$k$.
For convenience, we relax the restriction~\re{k>2n} and allow $k \ge 1$.

For the lowest values, $k = 1, 2, 3, 4$, the quantization conditions take the form
\begin{align}\notag\label{QCs}
 z^2 ={}& -4\sinh \mathcal F \sin \widetilde\Phi\,,
\\[2mm]\notag
z\ ={}& 4 \sinh \mathcal F \cos \widetilde\Phi\,,
\\[2mm]\notag
z^2={}& -6 \sinh \mathcal F \sin \widetilde\Phi -12 \sinh \mathcal F\sin (3 \widetilde\Phi)+6 \sinh (3 \mathcal F)
   \sin \widetilde\Phi
\\[2mm]\notag
{}&    
   +4 \sinh (3 \mathcal F) \sin (3 \widetilde\Phi)-6 \sqrt{3} \cosh \mathcal F
   \cos \widetilde\Phi+6 \sqrt{3} \cosh (3 \mathcal F) \cos \widetilde\Phi\,,
\\[2mm] 
z\ ={}& 4 \cosh (2\mathcal F)
   \sin (2 \widetilde\Phi)+4 \sinh (2\mathcal F)-4 \sin (2\widetilde\Phi) ,
\end{align}
where the notation was introduced for a modified phase
\begin{align}
\widetilde \Phi(z) = \Phi(z)-{\pi\over k}\,.
\end{align}
The phase $\widetilde{\Phi}(z)$ is introduced because, by virtue of~\re{Phi-par}, it is an odd function of~$z$.
Recall that the function $\mathcal{F}(z)$ shares the same property,  see \re{D-def}.
For odd and even values of $k$, the left-hand side of the relations~\re{QCs} is, respectively, even and odd in~$z$.
It is straightforward to verify that each term on the right-hand side of~\re{QCs} possesses a definite parity under the transformation $z \to -z$, matching that of the corresponding left-hand side.
  
The relations \re{QCs} take a particularly suggestive form.  
For higher values of~$k$, the right-hand side of the quantization condition can be expressed as a linear combination of terms of the type  
$\sinh(m_1 \mathcal{F}) \cos(m_2 \widetilde{\Phi})$ and $\cosh(m_1 \mathcal{F}) \sin(m_2 \widetilde{\Phi})$ (with $m_1, m_2$ positive integer) when $k$ is even,  
and of the type  
$\sinh(m_1 \mathcal{F}) \sin(m_2 \widetilde{\Phi})$ and $\cosh(m_1 \mathcal{F}) \cos(m_2 \widetilde{\Phi})$ when $k$ is odd.

Solving the quantization conditions \re{QCs}, we can determine the phase $\Phi(z)$ in terms of the function $\mathcal F(z)$. It is convenient to invert this dependence and express 
the function $D(z)=e^{\mathcal F(z)}$ in terms of the phase. For $k=1,2$ we find from \re{QCs}
\begin{align} 
D(z) {}&=\frac{1}{4\sin \Phi(z)}  \left(\sqrt{z^{4/k}+16\sin^2 \Phi(z)}+z^{2/k}\right).
\end{align}
For $z=0$ this relation correctly reproduces $D(0)=1$ (see \re{D-def}). At large positive $z$ we find instead
\begin{align}\label{D-large-z}
D(z) = {z^{2/k}\over 2\sin\Phi(z)} + O(z^{-2/k})\,.
\end{align}
It is straightforward to verify that the last two relations in \re{QCs} imply the same asymptotic behavior for $k=3,4$. Furthermore, using~\re{QC-e} and~\re{QC-o}, one can show that
the relation \re{D-large-z} holds for arbitrary $k$. 

\subsection{Potential}  

Before proceeding to the evaluation of the generating function~\re{W-psi}, we introduce the final ingredient of our method. 

We can generalize the definition of the function $E(x)$ in~\re{psi} by defining  
\begin{align}\label{En}
E_n(x) = e^{\frac{x}{2k}(2n+1)}\,,
\end{align}
where $n$ is a nonnegative integer. For $n=0$, this function reduces to  $E(x)$. 
Next, we define the matrix element of the resolvent of the operator~\re{rho-oper} with respect to the functions~\re{En} 
\begin{align}\label{Vn}
V_n(z) = z\VEV{E_n\left|\frac{1}{1+z \bm\rho}\bm\chi\right|E}
= z\VEV{E\left|\frac{1}{1+z \bm\rho}\bm\chi\right|E_n},
\end{align}
where the second equality follows from the representation $\bm\rho = \bm\chi\bm K / (4\pi k)$ and from the symmetry of the kernel \re{K-ker} of the operator 
$\bm K$ under the exchange of its arguments. The quantities~\re{Vn} are commonly referred to as \emph{potentials}, and they play a central role in the analysis of Fredholm determinants of integrable kernels.
 
For $n=0$, the potential $V_0(z)$ satisfies the differential equation
\begin{align}\label{dV0}
\partial_z V_0(z)
= \biggl\langle E\biggl|\frac{1}{(1+z\bm\rho)^2}\bm\chi\biggr|E\biggr\rangle
= \int_{-\infty}^{\infty}\! dx\, \chi(x)\, \psi^2(x|-z)\,.
\end{align} 
 Furthermore, for arbitrary $n$, the potential \re{Vn} can be expressed in terms of the $\psi-$functions defined in \re{psi}. 
To show this, we note that the potential \re{Vn} can be written through the function $\Gamma(x,y|z)$ defined in \re{Gamma-K}, evaluated at purely imaginary values of its arguments,
\begin{align}\label{V-Gamma} 
V_{n}(z) = 2\pi\, \Gamma(-i\pi,\, i\pi(2n+1)\,|\,z)\,.  
\end{align}
A derivation of this relation is presented in Appendix~\ref{app:C}. 

Substituting the representation of $\Gamma(x,y|z)$ from \re{R-psi1}, we obtain
\begin{align}\label{Vn0}
V_{n}(z)= \frac{2\pi z\, e^{- \mathcal F(z)}}{\sin(\frac{2\pi(n+1)}{k} )}
\Im \Big[ \psi(i\pi(2n+1)|z)\,\psi(i\pi|z)\Big].
\end{align}
This relation is a special case of the general formula \re{Vnm} derived in Appendix~\ref{app:C}. As explained above, the values of the $\psi-$function appearing in~\re{Vn0} can be expressed in terms of the real functions $r(z)$ and $\Phi(z)$ introduced in~\re{rho}, which, in turn, can be expressed through the function $\mathcal F(z)$ defined in \re{F-def}.
   
\subsection{Semiclassical approximation}

The relations for the function $\psi(x|z)$ derived in this and the previous sections hold for arbitrary values of $k$. Recall that large and small values of $k$ correspond, respectively, to the weak- and strong-coupling regimes of the ABJM theory. Our analysis shows that $\psi(x|z)$ as a function of $x$ has different analyticity properties in these two regimes.

At large $k$, the function $\psi(x|z)$ can be expanded perturbatively in powers of $1/k$, see \re{psi-weak} below. Each term in this expansion is polynomial in $x$ and $z$. This behavior is consistent with the expected analytical properties of  the function $\psi(x|z)$. Namely, for $k\to\infty$, the analyticity strip $-\pi k < \Im x < \pi k$ depicted in Figure~\ref{fig:poles} extends over the entire complex $x$-plane.

In contrast, for $k \to 0$, the analyticity strip collapses to the real $x$-axis. For real $x$, the function $\psi(x|z)$ satisfies the Baxter equation~\re{Bax}. Interpreting this equation as a Schr\"odinger-type equation with $k$ playing the role of the Planck constant, one may attempt a solution in the form of a WKB expansion
\begin{align}\label{psi-WKB}
\psi_{_{\rm WKB}}(x|z) = e^{\frac{1}{k} S_0(x|z) + S_1(x|z)  + O(k)} \,.
\end{align}
Substituting this ansatz into~\re{Bax} and matching terms order by order in $k$, leads to a hierarchy of equations for the functions $S_m(x|z)$ with $m=0,1,\dots$.
   
Trying to apply the semiclassical approximation to the computation of the generating function~\eqref{W-psi}, we encounter several obstacles. 
First, the Baxter equation does not uniquely determine the function $\psi(x|z)$. Indeed, the relation \re{Bax} is invariant under multiplication by an arbitrary $i\pi k-$periodic function.
To lift this ambiguity and select a unique solution of~\eqref{Bax}, one must supplement the Baxter equation with the analyticity conditions for $\psi(x|z)$ discussed above.

A second complication arises from the structure of relation~\eqref{W-psi}, which involves the function $\psi(x+2\pi i n|z)$ evaluated for real~$x$. For finite $k$ obeying the constraint~\eqref{k>2n}, the point $x+2\pi i n$ remains inside the analyticity strip, so the function $\psi(x+2\pi i n|z)$ is well defined. In the semiclassical limit $k\to 0$, however, this point moves outside the strip.\footnote{This issue is absent for $n=0$ and, in particular, does not affect the computation of the partition function.} Consequently, the semiclassical evaluation of the generating function~\eqref{W-psi} requires an analytic continuation of the WKB solution~\re{psi-WKB} into the complex $x$-plane. Such a continuation is inherently ambiguous unless the analytic structure of $\psi(x|z)$ is fixed a priori. 
The computation of the Wilson loop in~\cite{Klemm:2012ii} adopted a specific prescription for this analytic continuation. The discrepancy with the numerical results, described in the Introduction, suggests that this choice is not correct.

In the next section we show that the generating function~\eqref{W-psi} can in fact be computed for arbitrary~$k$, without invoking the semiclassical approximation. Most importantly, this can be achieved without constructing explicit solutions of the Baxter equation \re{Bax}. 

\section{Derivation of the Wilson loop}\label{sect5}

In this section, we apply the results derived previously to evaluate the generating function~\re{W-psi}. 
For this purpose, it is convenient to recast the relation~\re{W-psi} in the form
\begin{align}\label{W-simp}
\mathcal W_n(z,k)
= \frac{i z}{2\sin(2\pi n/k)}\,\frac{e^{i\pi n(n-1)/k}}{4\pi k}
\left[e^{-i\pi n/k}\, w_n(-z)\;-\; e^{i\pi n/k}\, w_n(z)\right],
\end{align}
where we introduced the notation for the function
\begin{align}\label{wn}
w_n(z)=\int_{-\infty}^{\infty} dx \, e^{(n-1)x/k}\,\chi(x)\,\psi(x|-z)\,\psi(x+2i\pi n\,|\,z)\,,
\end{align}
depending on the winding number $n$.

In the preceding sections, we established several nontrivial identities satisfied by the functions $\psi(x|\pm z)$. We show below that the function $w_n(z)$ possesses a remarkable feature: 
the integral in~\re{wn} can be expressed entirely in terms of the function~$\mathcal F(z)$, without requiring the explicit form of the functions $\psi(x|\pm z)$.

\subsection{Recurrence relations}

Let us evaluate the function \re{wn} for $n=0,1,2$ and then extend the result to general $n$.

\subsubsection*{Winding number $n=0$}

For $n=0$ the integral in \re{wn} coincides with the one appearing in \re{dF}. This allows us to identify $w_0(z)$ with the derivative of the function $\mathcal F(z)$ defined in \re{F-def}   
\begin{align}\label{w0} 
w_0(z)=\int_{-\infty}^\infty dx \, e^{-{x\over k}}\chi(x)\psi(x|-z)\psi(x|z)= 4\pi k \partial_z \mathcal{F}(z)\,.
\end{align}
Combining this relation with \re{W-simp}, we correctly reproduce the relation \re{dF-W}.

\subsubsection*{Winding number $n=1$}
 
For $n=1$, applying relation~\re{shift} with $m=1$ yields
\begin{align}\label{w1}
w_1(z) = e^{\frac{i\pi}{k}} \int_{-\infty}^{\infty} dx\, \chi(x)\, \psi^2(x|-z)
+ \frac{1}{2k}\, \psi(i\pi|z) \int_{-\infty}^{\infty} dx\, \chi(x)\, \Gamma(x,-i\pi|z)\, \psi(x|-z).
\end{align}
The integrals in the first and second terms coincide with those entering \re{dV0} and \re{dpsi}, respectively. Using these relations, we obtain
\begin{align}\label{w1-f}
w_1(z)
= e^{\frac{i\pi}{k}}\, \partial_z V_0(z)
- 2\pi\, \psi(i\pi|z)\, z\, \partial_z \psi(-i\pi|-z)\,.
\end{align}
Substituting the expression for the potential $V_0(z)$ from \re{Vn0} and applying \re{par}, we can express $w_1(z)$ in terms of the special values \re{rho}.
 
\subsubsection*{Winding number $n=2$}

For $n=2$, we apply the relation~\re{shift} with $m=2$ to obtain
\begin{align}\notag\label{w2}
w_2(z) {}& =e^{{2i\pi\over k}}\int_{-\infty}^\infty dx \,e^{{x\over k}} \chi(x)\psi(x|-z)\psi(x|z)
\\\notag
{}& +{1\over 2k}\psi(3i\pi|z)\int_{-\infty}^\infty dx \,e^{{x\over k}} \chi(x)\Gamma(x,-i\pi |-z) \psi(x|-z)
\\
{}& -{1\over 2k}\psi(i\pi|z)\int_{-\infty}^\infty dx \,e^{{x\over k}} \chi(x)\Gamma(x,-3i\pi |-z) \psi(x|-z)\,.
\end{align}
Compared with \re{w1}, the integrals in this relation involve an additional factor of $e^{x/k}$. 
This factor can be eliminated by applying the identity (see \re{GG} in Appendix~\ref{app:A})
\begin{align}\label{G-G}
e^{x/k}\, \Gamma(x,y|-z) = \Gamma(x,y|z)\, e^{y/k} - z\, \psi(x|-z)\, \psi(y|z) \,.
\end{align}
Using this identity, the last two integrals in \re{w2} can be evaluated with the help of \re{dpsi} and \re{dV0}:
\begin{align}
\int_{-\infty}^{\infty} dx\, e^{x/k}\, \chi(x)\, \Gamma(x,y|-z)\, \psi(x|-z)
= - 4\pi k\, e^{y/k}\, z\, \partial_z \psi(y|-z) - z\, \psi(y|z)\, \partial_z V_0(z)\,,
\end{align}
after substituting $y=-i\pi$ and $y=-3i\pi$. 

The integral in the first line of \re{w2} requires special attention. It can be recognized as a special case of the general integral $U_n(z)$, defined in \re{I,U} and \re{U-rec} below, evaluated at $n=2$.
In this way, we obtain
\begin{align}\notag\label{w2-f}
w_2(z)=e^{{2i\pi\over k}}U_2(z)
{}&+{1\over 2k}\psi(3i\pi|z)\lr{-4\pi k e^{-{i\pi\over k}} z \partial_z \psi(-i\pi|-z)-z\psi(-i\pi|z)\partial_z V_0(z)}
\\
{}&-{1\over 2k}\psi(i\pi|z)\lr{-4\pi k e^{-{3i\pi\over k}} z \partial_z \psi(-3i\pi|-z)-z\psi(-3i\pi|z)\partial_z V_0(z)},
\end{align}
where $U_2(z)$ is given by (see \re{U-rec})
\begin{align}
U_2(z) = \partial_z V_1(z) -{1\over 4\pi k} V_{0}(-z)\partial_z V_{0}(z)\,,
\end{align}
and  the potential $V_n(z)$ is defined in \re{Vn0}.
  
\subsubsection*{Arbitrary winding number}

For arbitrary positive integer $n$, we combine together \re{wn} and \re{shift} to find
\begin{align}\label{wn-sum}
w_n(z)=e^{{i\pi n\over k}} U_n(z)+ {1\over 2k} \sum_{m=0}^{n-1}(-1)^{m} 
  \psi(i\pi (2 (n-m)-1)|z) I_n(-i\pi(2m+1)|z)\,,
\end{align}
where the notation was introduced for two different types of integrals
\begin{align}\label{I,U}\notag
{}& U_n(z) = \int_{-\infty}^\infty dx \, e^{(n-1)x\over k}\chi(x)\psi(x|-z) \psi(x|(-1)^n z)\,,
\\[2mm]
{}& I_n(y|z) = \int_{-\infty}^\infty dx \, e^{(n-1)x\over k}\chi(x)\psi(x|-z)\Gamma(x,y|(-1)^{n+1}z)\,.
\end{align}

For $n=0$ and $n=1$ the relation \re{wn-sum} reduces to \re{w0} and \re{w1}. The corresponding values of the functions \re{I,U} can be identified as
\begin{align}\notag\label{bc}
{}& U_0(z)=4\pi k \partial_z \mathcal F(z)\,,\qquad U_1(z)=\partial_z V_{0}(z)\,, 
\\[2mm]
{}& I_1(y|z)=-4\pi kz\partial_z \psi(y|-z)\,.
\end{align} 
For $n\ge 1$, we can show that the functions \re{I,U} satisfy a coupled system of recurrence relations 
\begin{align}\label{U-rec}\notag
{}&U_n(z)= \partial_z V_{n-1}(z) -{1\over 4\pi k}\sum_{m=1}^{n-1} V_{n-m-1}((-1)^{m} z)U_{m}(z)\,.
\\[2mm]
{}&I_n(y|z)= e^{y/k}\, I_{n-1}(y|z) + (-1)^{n-1} z\, \psi(y|(-1)^n z)\, U_{n-1}(z)\,,
\end{align}
where the potentials $V_n(z)$ are defined in \re{Vn0}.
The second relation in \re{U-rec} can be obtained by applying the identity \re{G-G} and replacing $\Gamma(x,y|(-1)^{n+1} z)$ in \re{I,U} with its expression in terms of $\Gamma(x,y|(-1)^n z)$. The first and second terms on the right-hand side of this relation correspond to the respective terms in \re{G-G}.
The derivation of the first relation in \re{U-rec} can be found in Appendix~\ref{app:rr}. 
 
Being combined with the initial conditions \re{bc}, the first relation in \re{U-rec} allows us to determine $U_n(z)$ for arbitrary positive $n$. In particular, we verify that for $n=1$ it correctly reproduces the result \re{bc} for $U_1(z)$. Continuing the second relation in \re{U-rec}, we find 
\begin{align}\label{I-U}
I_n(y|z) = e^{y(n-1)\over k} I_1(y|z)+ z\sum_{m=1}^{n-1}(-1)^m e^{y(n-1-m)\over k} \psi(y|(-1)^{m+1}z) U_m(z)\,,
\end{align}
where $I_1(y|z)$ is given by \re{bc}. Substituting the explicit expressions for $U_n(z)$ into this relation, we can determine the functions $I_n(y|z)$ and, finally, compute the functions \re{wn-sum} for arbitrary $n$ entirely in terms of the special values $\psi(i\pi(2m+1)|\pm z)$ (with $m\in\mathbb{Z}$) and their derivatives. These values were computed in the previous section in terms of the functions $\mathcal F(z)$ and $\Phi(z)$.  Recall that these two functions are related to each other  through the quantization conditions \re{QCs}.
  
Taking into account the relations \re{rho}, \re{rho-F}, \re{psi3} and \re{Vn0}, we find from \re{wn-sum} that the resulting expressions for $w_n(z)$ are given by linear combinations of the derivatives $\partial_z \mathcal F(z)$ and $\partial_z \Phi(z)$ with the coefficient functions depending on $\mathcal F(z)$ and $\Phi(z)$. To save space we do not present their expressions. 

\subsection{Results}
  
Having determined the functions $w_n(z)$, we can now use the relation \re{W-simp} to evaluate the generating function $\mathcal W_n(z)$ for arbitrary positive integer $n$. 
We find that $\mathcal W_n(z)$ can be written in the form
\begin{align}\label{W-cf} 
\mathcal W_n(z,k)
= \frac{i^{\,n-1}}{\sin (\ft{2\pi n}{k} )}
\Big[
\mathcal A_n(z)
+ i\, z\, \partial_z \Phi(z)\, \mathcal B_n(z)
+ i\, z\, \partial_z \mathcal F(z)\, \mathcal C_n(z)
\Big],
\end{align}
where $\mathcal A_n(z)$, $\mathcal B_n(z)$, and $\mathcal C_n(z)$ are real functions of~$z$.  
Notice that the last two terms in the brackets are purely imaginary.

For the first few values of $n$, the coefficient functions in~\re{W-cf} take particularly compact forms:

\begin{itemize}

\item For $n=1$  
\begin{align}\notag\label{A-n=1}
{}& \mathcal A_1(z)=\sinh\mathcal F(z)\sin\Phi(z)\,,
\\[2mm]\notag
{}& \mathcal B_1(z)=-\sinh\mathcal  F(z) \left[2 \cot (\ft{2 \pi }{k})\cos \Phi (z)+\sin \Phi (z)\right]\,,
\\[2mm]
{}& \mathcal C_1(z)=-\sinh\mathcal  F(z) \cos \Phi (z)-\cosh\mathcal  F(z) \left[2\cot (\ft{2\pi
   }{k}) \sin \Phi (z)-\cos \Phi (z)\right]\,.
\end{align}

\item 
For $n=2$  
\begin{align}\notag\label{A-n=2}
 \mathcal A_2(z){}&=  (\cosh (2\mathcal  F(z))-1) \left(\cos^2 (\ft{2\pi }{k})-\cos (2 \Phi
   (z))\right)+\sinh (2\mathcal  F(z)) \sin ^2(\ft{2 \pi }{k}) \,,
\\[2mm]\notag
 \mathcal B_2(z){}&=(\cosh (2\mathcal  F(z))-1) \left[\cos (2 \Phi (z))-{2 \left(2 \cos (\ft{4 \pi }{k})+1\right)\over \sin (\frac{4 \pi }{k})}\sin (2 \Phi (z))-\ft{1}{2} (\cos(\ft{4
   \pi }{k})+3) \right]
\\{}&   \notag
+\sinh (2\mathcal  F(z))\cos ^2(\ft{2 \pi}{k}) \,,
\\[3mm]\notag
\mathcal C_2(z){}&=  - (\cosh (2\mathcal  F(z))-1) \left[\cos (\ft{4 \pi }{k}) \tan (\ft{2 \pi }{k})+ \sin (2 \Phi(z))\right]-
\cos (\ft{4 \pi }{k}) \tan (\ft{2 \pi }{k})
\\ {}&    
   +\sinh (2\mathcal  F(z)) \left[\sin (2 \Phi (z)) +{2 \left(2 \cos (\ft{4 \pi }{k})+1\right)\over \sin (\frac{4 \pi }{k})}\cos (2 \Phi
   (z))-\left(\cos (\ft{4 \pi }{k})+2\right) \cot (\ft{2 \pi}{k})\right].
\end{align}
\end{itemize}
We emphasize that these relations hold for arbitrary values of $k$ and $z$.  
The phase $\Phi(z)$ obeys the quantization conditions \re{QCs} and depends on both $k$ and the function $\mathcal F(z)$.

For higher values of $n$ the expressions for the coefficient functions become increasingly cumbersome and we refrain from displaying them here.~\footnote{Explicit expressions for the coefficient functions $\mathcal A_n$, $\mathcal B_n$ and $\mathcal C_n$ for $n\le 5$ can be found in an ancillary {\tt Mathematica} file attached to this submission.}  
Instead, we demonstrate below that the coefficient functions in \re{W-cf} acquire significantly simpler forms in the limits of small and large $z$.

Recall that $\mathcal W_n(z)$ is the generating function of the $1/6$ BPS Wilson loop.  
According to relation~\re{W-1/2}, the generating function of the $1/2$ BPS Wilson loop is expressed in terms of $\mathcal W_n(z)$ as
\begin{align}\label{calW-1/2}
\mathcal W^{1/2}_n(z)
= \mathcal W_n(z,k) - (-1)^n \,\overline{\mathcal W_n(z,k)}
= \frac{2 \, i^{n-1}}{\sin(\frac{2\pi n}{k})}\, \mathcal A_n(z)\, .
\end{align}
The last two terms inside the brackets in~\re{W-cf} do not contribute to $\mathcal W^{1/2}_n(z)$.  
We also note that the expressions for the coefficient function $\mathcal A_n(z)$ are substantially simpler than those for $\mathcal B_n(z)$ and $\mathcal C_n(z)$.  
This simplification can be viewed as a manifestation of the enhanced supersymmetry of the $1/2$ BPS Wilson loop.

To illustrate the remarkable simplicity of the coefficient functions $\mathcal A_n(z)$, let us examine the expression for $\mathcal A_1(z)$ at the specific value of the Chern–Simons level $k=4$. Using the last relation in~\re{QCs}, we find that the phase $\Phi(z)$ obeys
\begin{align}
\cos(2\Phi(z)) = {\sinh(2\mathcal F(z))-\frac14 z\over \cosh(2\mathcal F(z))-1}\,.
\end{align} 
Substituting this expression into~\re{A-n=1}, we obtain for $k=4$
\begin{align}\label{A_1-k=4}
\mathcal A_1 = \frac14\sqrt{z-4+4\, e^{-2\mathcal F(z)}}\,.
\end{align}
Combined with the relation \re{calW-1/2}, this leads to an exact expression for the generating function of the $1/2$ BPS Wilson loop in terms of the free energy $\mathcal F(z)$. We recall that the partition function $\Xi(z,k=4)$, and hence the free energy $\mathcal F(z)$, admits a closed-form representation in terms of Jacobi theta functions~\cite{Grassi:2014uua}. Together with relation~\re{A_1-k=4}, this yields an exact expression for the $1/2$ BPS Wilson loop at $k=4$.

In a similar manner, for arbitrary values of $k$ we can use the quantization conditions \re{QC-e} and \re{QC-o} to determine the phase $\Phi(z)$ in terms of $\mathcal F(z)$ and then apply the first relation in~\re{A-n=1} to derive an equation for the coefficient function $\mathcal A_1(z)$. For example, carrying out this procedure for $k=6$ leads to the polynomial equation
\begin{align}\notag\label{A_1-k=6}
\left(64 \mathcal A_1^3+24 \mathcal A_1-z\right)^2-9
\left(256 \mathcal A_1^4+96 \mathcal A_1^2-4 z \mathcal A_1+3\right)   e^{-2\mathcal  F(z)} 
\\[2mm]   
   +54 \left(8 \mathcal A_1^2+1\right) e^{-4\mathcal  F(z)}-27 e^{-6\mathcal  F(z)}=0\,.
\end{align} 

Moreover, using the first relation in~\re{A-n=1}, we can express the phase $\Phi(z)$ directly in terms of $\mathcal A_1(z)$. Substituting this expression into the remaining relations~\re{A-n=1} and~\re{A-n=2} then allows us to express all other coefficient functions solely in terms of $\mathcal A_1(z)$ and $\mathcal F(z)$. For instance,
\begin{align}\label{A2-A1}
\mathcal A_2(z) = 4\,\mathcal A_1^2(z) +\lr{1-e^{-2\mathcal F(z)}}\sin^2\!\Big(\frac{2\pi}{k}\Big)\,.
\end{align}
This relation is valid for arbitrary $k$ and $z$. 

Substituting \re{A_1-k=4} and \re{A2-A1} into \re{calW-1/2} yields the results \re{W-ex1} and \re{W-ex2} for the $1/2$ BPS Wilson loop.

\subsection{Weak coupling regime}  

As a nontrivial check of  the exact relations \re{W-cf} and \re{calW-1/2} for the generating functions of supersymmetric Wilson loops, we can verify that the large-$k$ expansion of $\mathcal W_n$ correctly reproduces the weak-coupling expansion of the Wilson loop.

At large $k$, the kernel~\eqref{rho-fun} scales as $O(1/k)$. This allows us to expand the matrix element in \re{psi} in powers of $\bm\rho$, or equivalently in powers of $z$, and thereby derive the $1/k$ expansion of the function $\psi(x|z)$,
\begin{align}\label{psi-weak}
\psi(x|z)=\sqrt{2}\,\bigg(1+\frac{2 x+z}{4 k}+\frac{2 x^2+z^2}{16\, k^2}+\frac{4 x^3-6 x^2 z+3 z^3}{192\, k^3}+O(1/k^4)\bigg).
\end{align}
Using this result, we obtain from \re{rho} and \re{rho-F} the corresponding large-$k$ expansions for the phase $\Phi(z)$ and the free energy $\mathcal F(z)$,
\begin{align}\notag
{}& \Phi(z) =\frac{\pi }{k}-\frac{\pi  z}{4 k^2}+\frac{\pi ^3 z}{24 k^4}+\frac{5 \pi ^3 z^3-128 \pi ^5 z}{3840 k^6}+O(1/k^8)\,,
\\[2mm]
{}& \mathcal F(z) = \frac{z}{2 k}+\frac{z^3}{96 k^3}+\frac{z^5-20 \pi ^2 z^3}{2560 k^5}+\frac{3 z^7-84 \pi ^2 z^5+1792 \pi ^4 z^3}{172032 k^7}+O(1/k^9)\,.
\end{align}
Substituting these expansions into \re{A-n=1} and \re{A-n=2} yields the large-$k$ behavior of the coefficient functions $\mathcal A_n$, $\mathcal B_n$, and $\mathcal C_n$, which in turn allows us to determine the generating function~\re{W-cf}.

We verified (see an ancillary {\tt Mathematica} file) that the resulting expression for $\mathcal W_n(z,k)$ is in perfect agreement with the expected large $k$ expansion
\begin{align}\notag\label{W-weak}
\mathcal W_n(z,k){}&=\frac{z}{4 k}+\frac{z \left(-z+4 i \pi  n^2\right)}{16 k^2}+\frac{z \left(-8 \pi ^2 n^4+4 \pi  n^2 (4 \pi -i z)+z^2\right)}{64 k^3}
\\[2mm]
{}&
+\frac{z \left(12 i \pi  n^2 z^2+24 \pi ^2
    (n^2-1 )^2 z-32 i \pi ^3  (n^2-6 ) n^4-3 z^3\right)}{768 k^4} + O(1/k^5)\,.
\end{align}
This expansion can be derived either by expanding \re{det2} in powers of $\bm\rho$ or by substituting \re{psi-weak} into \re{W-psi} and performing the integration.

\section{Large $z$ regime} \label{sect7}
 
In this section, we determine the leading asymptotic behavior of the generating function \re{W-cf} in the limit $z \to \infty$, with the parameter $k$ held fixed.  
To analyse this regime, it is convenient to use the relation \re{z-mu} and replace the fugacity parameter $z$ with the chemical potential $\mu$.

In the limit $\mu\to\infty$, the generating function $\mathcal W_n(\mu)$ is expected to take the form  \re{W-conj} and \re{W-dec}.  
Neglecting exponentially suppressed nonperturbative contributions, we obtain from \re{W-conj} and \re{W-pt} 
\begin{align}\label{W-exp}
\mathcal W_n(z,k)
 = \frac{i^{\,n-1}\, e^{2n\mu/k}}{k\, \sin \!\left(\ft{2\pi n}{k}\right)}
 \left[
 \frac{k}{4}
 + i\!\left(\frac{\mu}{\pi} + C_n(k)\right)
 + O\!\left(e^{-4\mu/k}\right)
 \right].
\end{align}
This expression originates from the $m=0$ term in the sum \re{W-conj}. The remaining terms with $m \neq 0$ are exponentially suppressed in the limit $\mu \to \infty$.  
As noted above, the coefficient $C_n(k)$ has been computed previously using different methods, leading to mutually inconsistent values \re{A-mismatch}.

Let us apply the relation \re{W-cf} to derive \re{W-exp} and determine the coefficient $C_n(k)$.  
A key simplification in the large--$\mu$ limit follows from relation~\re{D-large-z}, which yields
\begin{align}\label{F-sim}
\mathcal F(\mu) = {2\mu\over k}-\log (2\sin\Phi(\mu)) + O(e^{-{4\mu\over k}})\,.
\end{align}
Substituting this expression into \re{W-cf}, we obtain the simplified form
\begin{align}\label{W-cf1} 
\mathcal W_n(\mu,k)=\frac{i^{\,n-1}}{\sin (\ft{2\pi n}{k})}\Big[\mathcal A_n(\mu) + i\Big( \mathcal B_n(\mu) -\mathcal C_n(\mu) \cot\Phi(\mu) \Big)\Phi'(\mu) +{2i\over k} \mathcal C_n(\mu) 
\Big],
\end{align}
where $\Phi' =\partial_\mu \Phi(\mu)$.

\subsection{Coefficient functions at large $\mu$} 

As before, we compute the coefficient functions in \re{W-cf1} for $n=1$ and $n=2$, and then generalize the result to arbitrary $n$.

\paragraph{$\boldsymbol{n=1}$:}
Substituting relation~\re{F-sim} into~\re{A-n=1}, we find that the coefficient functions simplify to
\begin{align}\notag
\mathcal A_1 &= \frac{1}{4}\, e^{2\mu/k} + \dots\,,
\\\notag
\mathcal B_1 &= -\frac{1}{4}\, e^{2\mu/k}
 \left( 2 \cot\Phi(\mu)\, \cot\!\left(\ft{2\pi}{k}\right) + 1 \right)
 + \dots\,,
\\
\mathcal C_1 &= -\frac{1}{2}\, e^{2\mu/k}\,
 \cot\!\left(\ft{2\pi}{k}\right) + \dots\,,
\end{align}
where the dots denote subleading corrections suppressed by a factor of
$e^{-4\mu/k}$.  
For $k=4$ and $k=6$, the resulting expression for $\mathcal A_1$ verifies the relations~\re{A_1-k=4} and~\re{A_1-k=6}, providing a nontrivial consistency check.

Combining these relations we obtain from \re{W-cf1}
\begin{align}\mathcal W_1(z,k)=  \frac{ e^{2\mu/k}}{k\, \sin (\frac{2\pi}{k} )}
  \left[{k\over 4}-i\lr{{k\over 4}  \Phi '(\mu)+\cot\Big(\frac{2 \pi }{k} \Big)}+ O(e^{-{4\mu\over k}})\right]  .
\end{align}
 
\paragraph{$\bm{n=2}$:} 
Repeating the same analysis for the coefficient functions in \re{A-n=2}, we find
\begin{align}\notag
{}& \mathcal A_2=\frac{1}{4} e^{\frac{4 \mu }{k}}+ \dots\,,
\\[1.2mm]\notag
{}& \mathcal B_2=-\frac14 e^{\frac{4 \mu }{k}}\left[2 \cot(\Phi(\mu))\lr{\cot  (\ft{2 \pi }{k} )+\cot  (\ft{4 \pi }{k} )}+1\right]+\dots\,,
\\[1.2mm]
{}& \mathcal C_2=-\frac12e^{\frac{4 \mu }{k}} \lr{\cot  (\ft{2 \pi }{k} )+\cot  (\ft{4 \pi }{k} )}+\dots\,.
\end{align}
where, as before, the dots denote corrections exponentially suppressed as $e^{-4\mu/k}$.
We verify that the  expressions for $\mathcal A_1(z)$ and $\mathcal A_2(z)$ satisfy the relation \re{A2-A1}. 

Substituting the above relations into \re{W-cf} we get
\begin{align}
\mathcal W_2(z,k)=  \frac{i\, e^{4\mu/k}}{k\, \sin (\frac{4\pi}{k} )}
  \left[{k\over 4}-i\lr{{k\over 4}  \Phi '(\mu)+\cot \left(\frac{2 \pi }{k}\right)+\cot \left(\frac{4 \pi }{k}\right)}+ O(e^{-{4\mu\over k}})\right].  
\end{align}

\paragraph{$\bm{n\ge 3}$:} The above expressions display a clear structural pattern, indicating that in the large-$z$ limit the coefficient functions obey
\begin{align}\notag
{}& \mathcal A_n =-\mathcal B_n+ \mathcal C_n \cot\Phi =\frac{1}{4} e^{\frac{2n \mu }{k}}+ \dots\,,
\\
{}& \mathcal C_n=-\frac12 e^{\frac{2n \mu }{k}}\sum_{m=1}^n \cot\! \Big(\frac{2 \pi m}{k}\Big)+\dots \,.
\end{align}
Further computations confirm that this structure persists for larger $n$. In particular, evaluating the coefficient functions for $3 \le n \le 10$ we found
\begin{align}\label{W-pre-fin}
\mathcal W_n(z,k)= \frac{i^{\,n-1}\, e^{2n\mu/k}}{k\, \sin \!\left(\ft{2\pi n}{k}\right)}\left[{k\over 4}-{ik\over 4}  \Phi '(\mu)-i\sum_{m=1}^n \cot\! \Big(\frac{2 \pi m}{k}\Big)+O(e^{-{4\mu\over k}})\right].  
\end{align}
The next step is to determine the large-$\mu$ behavior of the derivative of the phase $\Phi'(\mu)$.

\subsection{Phase at large $\mu$}

To determine the phase $\Phi(\mu)$, we complement relation~\re{F-sim} with the large-$\mu$ behavior of the free energy $\mathcal F(z)$.  
Using the definitions~\re{F-def} and~\re{D-def}, together with the expression~\re{Xi-conj} for the generating function $\Xi(z,k)$, we obtain
\begin{align}
e^{\mathcal F(\mu)}
 = \frac{\Xi(\mu,k)}{\Xi(\mu+i\pi,k)}
 = 
 \frac{\displaystyle \sum_{m=-\infty}^{\infty}
        e^{\,J(\mu + 2\pi i m,\,k)}}
      {\displaystyle \sum_{m=-\infty}^{\infty}
        e^{\,J(\mu + \pi i (2m-1),\,k)}}\,.
\end{align}
Neglecting exponentially suppressed contributions to $\mathcal F(\mu)$, we may replace $J(\mu)$ in this expression by its perturbative part~\re{J-pt}.  
In addition, only the $m=0$ term in the numerator and the $m=0$ and $m=1$ terms in the denominator need to be retained.

In this way, we arrive at
\begin{align}\notag
e^{\mathcal F(\mu)} {}&= {e^{J^{\text{pert}}(\mu , k)}\over  e^{J^{\text{pert}}(\mu - i\pi), k)}+e^{J^{\text{pert}}(\mu+i\pi), k)}}+O(e^{-{4\mu\over k}})
\\
{}&= {e^{2\mu/k}\over 2 \cos \Big(\displaystyle \frac{2}{\pi  k}\mu^2+\frac{\pi  k}{24}-\frac{\pi }{3 k}\Big)}+O(e^{-{4\mu\over k}})\,,
\end{align}
where, in the second step, we replaced $J^{\rm pert}(\mu)$ with its expression \re{J-pt}.
Comparing this relation with~\re{D-large-z} yields the large-$\mu$ behavior of the phase,
\begin{align}
\Phi(\mu) =-\frac{2}{\pi  k}\mu^2-\frac{\pi  k}{24}+{\pi\over 2}+\frac{\pi }{3 k}+ O(e^{-{4\mu\over k}})\,.
\end{align}
Substituting this expression into \re{W-pre-fin}, we reproduce \re{W-exp} and determine the coefficient $C_n=2e^{-\frac{2n \mu }{k}}\mathcal C_n$
\begin{align}
C_n(k)=-\sum_{m=1}^n \cot\! \Big(\frac{2 \pi m}{k}\Big),
\end{align}
in complete agreement with the numerical results in~\re{A-mismatch}.

\section{Concluding remarks}\label{sect8}

The localization reduces the computation of supersymmetric Wilson loops in the ABJM theory to finite-dimensional matrix integrals. In this paper, we develop new techniques for evaluating these integrals for arbitrary values of the parameters. Our approach is based on an operator representation of the Wilson loops within the Fermi gas formalism  in terms of the resolvent of the integral operator that previously appeared in the computation of the partition function of the ABJM theory on round sphere $S^3$.
 
We derived a set of nontrivial relations for the resolvent and used them to obtain exact expressions for the generating functions of Wilson loops in terms of the free energy. We showed that, in the large-$k$ limit at fixed $N$, these expressions reproduce the weak-coupling expansion of the Wilson loops, while in the large-$N$ limit at fixed $k$, they agree with high-precision numerical results. As a byproduct, our analysis clarifies the origin of the previously observed discrepancy between numerical data and the semiclassical expression for the $1/6$ BPS Wilson loop.

In the above analysis, we have neglected the exponentially small corrections to~\re{J-grand} and~\re{W-exp} at large~$N$. For large $k$ with fixed ’t~Hooft coupling $\lambda = N/k$, in the type IIA string theory regime, the nonperturbative contribution~\re{Jnp} arises from string worldsheet instantons wrapping the $\mathbb{CP}^{1}$ inside $\mathbb{CP}^{3}$~\cite{Drukker:2010nc,Cagnazzo:2009zh}, as well as from D2-brane instantons wrapping the $\mathbb{RP}^{3}$ cycle in $\mathbb{CP}^{3}$~\cite{Drukker:2011zy,Drukker:2010nc}. In the M-theory regime, corresponding to large $N$ at fixed $k$, these two types of corrections uplift to distinct M2-brane instantons wrapping, respectively, $S^{3}/\mathbb{Z}_{k}$ (including the 11-dimensional circle) and $\mathbb{RP}^{3} \subset \mathbb{CP}^{3}$ inside $S^{7}/\mathbb{Z}_{k}$ (not including the 11-dimensional circle) \cite{Cagnazzo:2009zh,Giombi:2023vzu,Gautason:2023igo,Beccaria:2023ujc}.~\footnote{We are grateful to Arkady Tseytlin for useful discussions regarding this point.}
  
The nonperturbative corrections in ABJM theory were systematically investigated using the refined topological string representation in~\cite{Nosaka:2015iiw,Hatsuda:2012dt,Hatsuda:2012hm,Hatsuda:2013oxa}, and through direct computations of the contribution of quantum M2-branes in~\cite{Giombi:2023vzu,Gautason:2023igo,Beccaria:2023ujc}.
High-precision numerical analyses in these works 
uncovered the intricate structure of instanton corrections to both the ABJM partition function~\cite{Hatsuda:2015gca} and supersymmetric Wilson loops~\cite{Hatsuda:2013yua,Okuyama:2016deu}. The method developed in this paper provides a way to understand this structure directly within the Fermi gas formalism, without relying on a conjectured duality with topological string theory~\cite{BKT2}.

As mentioned above, the matrix model integral \eqref{Z} obtained via localization is well-defined only for levels $ k $ satisfying \eqref{k>2n}. For $ k = 2n $, the integral develops a pole. This pole does not appear in the weak-coupling expansion \eqref{W-weak}, but it arises for fixed $ k $ due to the factor $ 1/\sin(2\pi n/k) $ in \eqref{W-exp}.  
In the large-$ N $ limit with fixed $ k $, the $ 1/2 $ BPS Wilson loop admits a dual description in terms of an M2-brane wrapping the M-theory circle. It was shown in~\cite{Giombi:2023vzu} that for $ n=1 $ and $ k>2 $, the same factor $ 1/\sin(2\pi/k) $ is precisely reproduced by the one-loop contribution in the partition function of the wrapped M2-brane. Interestingly, for $ k=1 $ and $ k=2 $, the holographic prediction for the Wilson loop remains finite, while the localization result becomes singular. This discrepancy calls for an explanation.

It would be interesting to generalize the method described in this paper to compute the latitude Wilson loop~\footnote{The matrix model for the latitude Wilson loop was proposed in \cite{Bianchi:2018bke} and tested at weak coupling up to three loops, both through Feynman-diagrammatic computations and via comparison with the bremsstrahlung function, showing nontrivial agreement. It was later derived in \cite{Griguolo:2021rke}, where it was shown, under mild assumptions, that supersymmetric localization on 
$S^3$ reproduces the expected matrix integral.} in ABJM theory~\cite{Griguolo:2021rke,Bianchi:2018bke}, as well as Wilson loops in the presence of deformations,  such as real masses \cite{Armanini:2024kww,Nosaka:2015iiw} and the squashing of the $S^3$ sphere, that break conformal invariance but preserve part of the supersymmetry. Notably, it has been conjectured (see \cite{Bobev:2025ltz} and references therein) that the large $N$ partition function can still be expressed in terms of an Airy function, suggesting that Wilson loops may have the same property.
 
\section*{Acknowledgements} 

We would like to thank Luca Griguolo, Luigi Guerrini, Shota Komatsu, Valentin Reys and Arkady Tseytlin for interesting discussions. We are indebted to 
Kazumi Okayama for sharing with us the results of his work \cite{Okuyama:2016deu}. We are also grateful to Marcos Mari\~no for useful comments.

The research of B.B. was supported by the Doctoral Excellence Fellowship Programme funded by the National
Research Development and Innovation Fund of the Ministry of Culture and Innovation and the Budapest
University of Technology and Economics, under a grant agreement with the National Research, Development
and Innovation Office (NKFIH). The work of G.K. and A.T.  was supported by the French National Agency for Research grant ``Observables'' (ANR-24-CE31-7996).
  
\appendix

\section{Derivation of \re{W-psi}}\label{app:A}
 
In this appendix, we present the derivation of relation \re{W-psi}.  

The relation~\re{det3} involves the function~\re{Gamma}, evaluated at $y = x + 2\pi i n$,  
\begin{align}\label{app:calW}
\mathcal{W}_n(z,k) = \int_{-\infty}^\infty dx\, e^{n( x+i\pi n)\over k}\frac{\chi(x)}{4\pi k}\, \Gamma(x, x + 2\pi i n \,|\, z) \,.
\end{align}
This function admits an equivalent representation~\re{Gamma-K} in terms of the operator~$\bm K$ defined in~\re{K-ker}.  
The kernel of this operator satisfies the relation
\begin{align}
K(x,y)\, e^{y/k} + e^{x/k} K(x,y) = E(x) E(y)\,,
\end{align}
where the function $E(x)$ is defined in~\re{psi}.  
Multiplying both sides of this relation by $\chi(x)/(4\pi k)$, we can promote it to the operator identity
\begin{align}
\bm\rho\, e^{\bm x/k} + e^{\bm x/k}\bm\rho = \frac{1}{4\pi k}\, \bm\chi \ket{E}\bra{E}\,,
\end{align}
where $\bm\chi = \chi(\bm x)$ and $\langle x | \bm\rho | y \rangle = \chi(x)\, K(x,y)/(4\pi k)$.
We can use this relation to get
\begin{align}\label{anti-com}
{z\bm\rho\over 1+ z\bm\rho}e^{\bm x/k}+e^{\bm x/k} {z\bm\rho\over 1- z\bm\rho}={z\over 4\pi k} {1\over 1+ z\bm\rho}\bm\chi \ket{E}
\bra{E}{1\over 1- z\bm\rho}\,.
\end{align}
Taking matrix elements of both sides and using \re{psi}, \re{Gamma}, and \re{app:id}, we find
\begin{align}\label{GG}
\Gamma(x,y|z) e^{y/k}-e^{x/k} \Gamma(x,y|-z)=z\chi(x) \psi(x|-z)\psi(y|z)\,.
\end{align}
Replacing $z\to -z$ in this relation and combining the result with \re{GG}, we obtain a system of linear equations for $\Gamma(x,y|z)$ and $\Gamma(x,y|-z)$. Solving this system yields the expression \re{R-psi}. Substituting \re{R-psi} into \re{app:calW}, we finally arrive at the relation \re{W-psi}.

\section{Parity properties}\label{app:B}

In this appendix, we derive the parity relation~\re{par}. 

Let us define a parity operator $\bm P$ by $\bm P\ket{x}=\ket{-x}$.
Using definition~\re{psi}, the function $\psi(-x|z)$ can then be written as
\begin{align}
\psi(-x|z)=\vev{E|{1\over 1-z\bm\rho}\bm P|x}=\vev{E|\bm P{1\over 1-z\bm\rho}|x}=\vev{E|e^{-{\bm x\over k}}{1\over 1-z\bm\rho}|x}\,.
\end{align}
In the first step, we used the fact that the operator~\re{rho-oper} preserves the parity, $[\bm{P}, \bm{\rho}] = 0$. 
In the last step, we employed the property of the function $E(x)$ defined in~\re{psi}, namely $E(-x) = e^{-\frac{x}{k}} E(x)$.

We employ the operator identity (see \re{anti-com})
\begin{align}
e^{-{\bm x\over k}}{1\over 1-z\bm\rho}-{1\over 1+z\bm\rho}e^{-{\bm x\over k}}={z\over 4\pi k} e^{-{\bm x\over k}}{1\over 1-z\bm\rho}\bm\chi\ket{E}\bra{E}{1\over 1+z\bm\rho}e^{-{\bm x\over k}}
\end{align}
to continue the previous relation as
\begin{align}\notag\label{f-fun}
{}&\psi(-x|z)=e^{-{x\over k}}\psi(x|-z)f(z)\,,
\\[1.5mm]
{}& f(z)= 1+{z\over 4\pi k}\bra{E}\bm P {1\over 1-z\bm\rho}\bm\chi\ket{E}\,.
\end{align}
We next show that the function $f(z)$ coincides with the ratio of the Fredholm determinants defined in \re{D-def}.

Differentiating both sides of \re{f-fun}, we obtain
\begin{align}\notag\label{df} 
\partial_z f(z) {}&={1\over 4\pi k}\int_{-\infty}^\infty dx\, \bra{E}\bm P {1\over 1-z\bm\rho} \ket{x}\bra{x}{1\over 1-z\bm\rho} 
\bm\chi\ket{E}
\\\notag
{}&={1\over 4\pi k}\int_{-\infty}^\infty dx\, \chi(x)\psi(-x|z)\psi(x|z)
\\ 
{}&={1\over 4\pi k}f(z) \int_{-\infty}^\infty dx\, \chi(x) e^{-{x\over k}}\psi(x|-z) \psi(x|z)\,,
\end{align}
where in the last relation we applied \re{f-fun}. In the second relation, we used the definition~\re{psi} and the identity
\begin{align}\label{app:id}
\bra{x}{1\over 1-z\bm\rho} \bm\chi\ket{E}=\bra{E} \bm\chi{1\over 1-z\bm\rho^t}\ket{x}
=\bra{E}{1\over 1-z\bm\rho} \bm\chi\ket{x}=\chi(x) \psi(x|z)\,,
\end{align}
where $\bm{\rho}^t = \bm{\chi}^{-1} \bm{\rho} \bm{\chi}$ denotes the transposed integral operator with kernel $\rho(y,x)$. 
 
 Comparing the relation \re{df} with \re{dF}, we conclude that $\partial_z \log f(z) = \partial_z \mathcal{F}(z)$.
Since both $\log f(z)$ and $\mathcal{F}(z)$ vanish at $z=0$ (cf. \re{f-fun} and \re{F-def}), it follows that
\begin{align}
f(z) = e^{\mathcal{F}(z)} = D(z)\,.
\end{align}
Combining this result with \re{f-fun}, we arrive at the parity relation~\re{par}.
 
\section{Properties of the potentials} \label{app:C}
  
In general, the potential is defined as the matrix element of the resolvent of the operator~\re{rho-oper} with respect to the functions introduced in~\re{En}
\begin{align}\label{Vnm-d}
V_{nm}(z) = V_{mn}(z)
= z\,\biggl\langle E_n\biggl|\frac{1}{1 + z\bm\rho}\,\bm\chi\biggr|E_m\biggr\rangle\,.
\end{align}
This is a real function of~$z$ that is symmetric under the exchange of indices~$n$ and~$m$.  
The symmetry property follows from the relation between the operator~$\bm\rho$ and its transpose,  
$\bm\chi\,\bm\rho^{\rm t} = \bm\rho\,\bm\chi$. In the special case of $m=0$, the potential $V_{n0}$ coincides with
the function $V_n(z)$ introduced in \re{Vn}.  

The potentials $V_{nm}(z)$ are not independent. 
Taking matrix elements of both sides of \re{anti-com} over the states $E_n$ and $E_m$, we obtain~\footnote{Similar relations have previously appeared in \cite{Tracy_1996}.}
\begin{align} \label{VV}
{}& V_{n,m+1}(-z) + V_{n+1,m}(z)={1\over 4\pi k}V_{n0}(-z)V_{0m}(z)\,.
\end{align}
These relations allow us to express $V_{nm}(z)$ for $m\neq 0$ in terms of $V_{n0}(z)$.  
 
 Furthermore, the potentials \re{Vnm-d} can be expressed in terms of the $\psi-$function introduced in \re{psi}.
This follows from the relation between the functions \re{Vnm-d} and \re{Gamma-K}
\begin{align}\notag\label{Vnm-id}
V_{nm}(z) {}&= 2\pi \Gamma(i\pi(2n+1),-i\pi(2m+1)|z)
\\[1.5mm]
{}&= 2\pi \Gamma(-i\pi(2n+1),i\pi(2m+1)|z)\,,
\end{align}
where the second equality is obtained from the first one by complex conjugation. 
Using the representation \re{R-psi1} of the $\Gamma-$function, we then find
\begin{align}\label{Vnm} 
V_{nm}(z)={2\pi z\, e^{-\mathcal F(z)}\over \sin({2\pi(n+m+1)\over k})}{}\Im\Big[\psi(i\pi(2n+1)|z)\psi(i\pi(2m+1)|z)  \Big].
\end{align}
This relation is well defined for $n+m+1<k/2$.

Let us now prove the relation \re{Vnm-id}. Substituting  \re{Vnm} and \re{Gamma-K} into \re{Vnm-id}, we arrive at the identity   
\begin{align} \label{id-M}
 \VEV{E_m\left|{1\over 1+z\bm\rho}\bm\chi\right|E_n} =2\pi\vev{i\pi(2m+1))|\bm K{1\over 1+z \bm\rho}|-i\pi(2n+1)}\,,
\end{align}
where $E_n(x) = e^{x(2n + 1)/(2k)}$ and $\bm{\chi} = 1/(2\cosh(\bm{x}/2))$. The operators $\bm{\rho}$ and $\bm{K}$ are defined in~\re{rho-fun} and~\re{K-ker}, respectively, and are related by $\bm\rho=\bm\chi\bm K/(4\pi k)$.

Expanding both sides of \re{id-M} in powers of $z$, we obtain
\begin{align} \label{id-M1}
 \VEV{E_m\left|\bm\chi(\bm K\bm\chi)^\ell\right|E_n}=2\pi\vev{i\pi(2m+1))|(\bm K\bm\chi)^\ell\bm K|-i\pi(2n+1)}\,.
\end{align}
For $\ell=0$, this relation follows directly from the integral representation of the kernel~\re{K-ker},
\begin{align}\label{K-chi}
\vev{x|\bm K|y} = \int_{-\infty}^\infty {dp\, e^{-ip(x-y)/(2\pi k)}\over 2\pi \cosh(p/2)} =  \int_{-\infty}^\infty {dp\over\pi}\, e^{-ip(x-y)/(2\pi k)}\chi(p)\,.
\end{align}
Indeed, substituting $x = i\pi(2m + 1)$ and $y = -i\pi(2n + 1)$ into this relation, the integral on the right-hand side can be identified with  $\int_{-\infty}^\infty dp\, E_m(p) \chi(p) E_n(p)/(2\pi)$, in agreement with \re{id-M1}. Note that the integral in~\re{K-chi} converges for $-\pi k < \Im(x - y) < \pi k$, and therefore the matrix elements $\langle E_m|\bm{\chi}|E_n\rangle$ are well-defined provided $m + n < k/2 - 1$.
   
The proof of~\re{id-M1} relies on the following identity
\begin{align}\notag\label{inter}
\bra{i\pi(2m+1)}\bm K\bm\chi\bm K\ket{x}
{}&= \int_{-\infty}^\infty \frac{dp}{\sqrt{2}\pi}\, E_m(p)\chi(p)
     \int_{-\infty}^\infty dy\, e^{ipy/(2\pi k)}\chi(y) K(y,x)
\\\notag
{}&=\int_{-\infty}^\infty \frac{dp}{\sqrt{2}\pi}\, E_m(p)\chi(p)
     \int_{-\infty}^\infty dy\, e^{ipy/(2\pi k)}\chi(y)
     \int_{-\infty}^\infty \frac{dp'}{\pi}\, e^{ip'(x-y)/(2\pi k)}\chi(p')
\\
{}&=\frac{1}{\sqrt{2}\pi}\int_{-\infty}^\infty dp\,dp'\, E_m(p)\chi(p)
     K(p,p')\chi(p')\, e^{ip'x/(2\pi k)}\,.
\end{align}
In the first step, we used relation~\re{K-chi} for $x=i\pi(2m+1)$ and expressed the $m$–dependent exponential factor in terms of the function $E_m(p)$.  
In the second step, the kernel $K(y,x)$ was replaced by its representation~\re{K-chi}.  
Finally, in the last step, we applied~\re{K-chi} once more to evaluate the $y$–integral  in terms of $K(p,p')$.

Substituting $x=-i\pi(2n+1)$ in~\re{inter} and identifying the $n$–dependent exponential factor as $E_n(p')/\sqrt{2}$, we obtain identity~\re{id-M1} for $\ell=1$.  
Applying the operator $\bm{\chi}\bm{K}$ successively $(\ell - 1)$ times to both sides of~\re{inter} and repeating the same steps yields the general identity~\re{id-M1}, and consequently, the relation~\re{id-M}.  
  
\section{Recurrence relation}\label{app:rr}

In this appendix, we derive the recurrence relations \re{U-rec} for the functions $U_n(z)$ introduced in \re{I,U}.

We begin by examining the derivative of the potential $\partial_z V_{n-1}(z)$, which appears on the right-hand side of \re{U-rec}. In close analogy with \re{dV0}, we obtain
\begin{align}\notag 
\partial_z V_{n-1}(z)
{}& = \biggl\langle E_{n-1}\biggl|\frac{1}{(1+z\bm\rho)^2}\bm\chi\biggr|E\biggr\rangle
\\
{}& = \int_{-\infty}^{\infty}\! dx\, \chi(x)\, \psi(x|-z)\VEV{E|e^{(n-1)\bm x/k} {1\over 1+z\bm \rho}|x},
\end{align} 
where in the second line we used \re{En}. Combining this relation with \re{I,U}, we find
\begin{align}\label{U-dV}
U_n(z) - \partial_z V_{n-1}(z) = \int_{-\infty}^{\infty}\! dx\, \chi(x)\, \psi(x|-z)\VEV{E\bigg|{1\over 1-(-1)^n z\bm \rho}e^{(n-1)\bm x/k} -e^{(n-1)\bm x/k} {1\over 1+z\bm \rho}\bigg|x}.
\end{align}
To simplify the matrix element on the right-hand side, we apply the identity \re{anti-com}. This gives
\begin{align}\notag\label{me}
\bra{E} e^{(n-1)\bm x/k} {1\over 1+z\bm \rho}\ket{x} {}& =  \bra{E}e^{(n-2)\bm x/k}\left[{1\over 1-z\bm \rho}e^{\bm x/k} -{z\over 4\pi k}{1\over 1-z\bm \rho}\bm\chi \ket{E}\bra{E} {1\over 1+z\bm \rho}\right]\ket{x}
 \\
{}& =\bra{E}e^{(n-2)\bm x/k}{1\over 1-z\bm \rho}\ket{x} e^{x/k}+{1\over 4\pi k}V_{n-2}(-z)\psi(x|-z)
\end{align}
where in the second step we used \re{Vn} together with \re{psi}. Iterating this identity repeatedly, we obtain the following expression for the matrix element  on the left-hand side of \re{me}  
\begin{align}
 \bra{E}{1\over 1-(-1)^n z\bm \rho}e^{(n-1)\bm x/k}\ket{x}+ {1\over 4\pi k}\sum_{m=1}^{n-1} V_{n-m-1}((-1)^{m} z)\psi(x|(-1)^{m} z)\,.
\end{align}
Substituting this expression into \re{U-dV} yields the recurrence relation \re{U-rec}.

\bibliographystyle{JHEP}    
\bibliography{BKT}
     
\end{document}